\documentclass{aastex}
\usepackage{epsfig}
\usepackage{amsmath}
\usepackage{mathrsfs}
\usepackage{subfigure}
\usepackage{textcomp}
\usepackage[normalem]{ulem}

\newcommand{\begeq}{\begin{equation}}
\newcommand{\fineq}{\end{equation}}
\newcommand{\begeqarray}{\begin{eqnarray}}
\newcommand{\fineqarray}{\end{eqnarray}}

\newcommand{\lapprox}{\lower.4ex\hbox{$\;\buildrel <\over{\scriptstyle\sim}\;$}}
\newcommand{\gapprox}{\lower.4ex\hbox{$\;\buildrel <\over{\scriptstyle\sim}\;$}}
\newcommand{\Mach}{\mathscr{M}}

\slugcomment{accepted for publication in ApJ}
\shorttitle{Gas and Radiation Pressure Effects on Accretion Column Properties}
\shortauthors{West, Wolfram, \& Becker}

\begin{document}

\title{DYNAMICAL AND RADIATIVE PROPERTIES OF X-RAY PULSAR ACCRETION COLUMNS: PHASE-AVERAGED SPECTRA}
\author{Brent F. West}
\affil{Unites States Naval Academy, Annapolis, MD; bwest@usna.edu}
\author{Kenneth D. Wolfram}
\affil{Naval Research Laboratory (retired), Washington, DC; kswolfram@gmail.com}
\author{Peter A. Becker}
\affil{Department of Physics and Astronomy, George Mason University, Fairfax, VA USA; pbecker@gmu.edu}

\begin{abstract}
The availability of the unprecedented spectral resolution provided by modern X-ray observatories is opening up new areas for study involving the coupled formation of the continuum emission and the cyclotron absorption features in accretion-powered X-ray pulsar spectra. Previous research focusing on the dynamics and the associated formation of the observed spectra has largely been confined to the single-fluid model, in which the super-Eddington luminosity inside the column decelerates the flow to rest at the stellar surface, while the dynamical effect of gas pressure is ignored. In a companion paper, we have presented a detailed analysis of the hydrodynamic and thermodynamic structure of the accretion column obtained using a new self-consistent model that includes the effects of both gas and radiation pressure. In this paper, we explore the formation of the associated X-ray spectra using a rigorous photon transport equation that is consistent with the hydrodynamic and thermodynamic structure of the column. We use the new model to obtain phase-averaged spectra and partially-occulted spectra for Her X-1, Cen X-3, and LMC X-4. We also use the new model to constrain the emission geometry, and compare the resulting parameters with those obtained using previously published models. Our model sheds new light on the structure of the column, the relationship between the ionized gas and the photons, the competition between diffusive and advective transport, and the magnitude of the energy-averaged cyclotron scattering cross section.

\end{abstract}

% The different journals have different requirements for keywords. The
% keywords.apj file, found on aas.org in the pubs/aastex-misc directory,
% contains a list of keywords used with the ApJ and Letters. These are
% usually assigned by the editor, but authors may include them in their
% manuscripts if they wish.

\section{INTRODUCTION}

Nearly a half-century has passed since the first observations of X-ray emission from accretion-powered X-ray pulsars. In these sources, the gravitational potential energy of the accreting gas is efficiently converted into kinetic energy in an accretion column that is collimated by the strong magnetic field ($B \sim 10^{12}\,$G). Since the surface of the neutron star represents a nearly impenetrable barrier, the kinetic energy is converted into thermal energy at the bottom of the column, and escapes through the column walls in the form of X-rays. Hence the emergent X-ray luminosity, $L_{\rm X}$, is essentially equal to the accretion luminosity, $L_{\rm acc} \equiv G M_* \dot M/R_*$, where $M_*$ and $R_*$ denote the stellar mass and radius, respectively, and $\dot M$ is the accretion rate. The X-ray emission is generated along the column and in the vicinity of one or both of the magnetic polar caps, thereby forming ``hot spots'' on the stellar surface. The X-rays were initially thought to emerge as fan-shaped beams near the base of the accretion column, but it soon became clear that a pencil beam component was sometimes necessary in order to obtain adequate agreement with the observed pulse profiles (Tsuruta \& Rees 1974; Bisnovatyi-Kogan \& Komberg 1976; Tsuruta 1975).

Analysis of X-ray pulsar spectra is usually performed by averaging over many rotation periods to obtain phase-averaged profiles. The resulting spectra are characterized by a power-law continuum extending up to a high-energy exponential (thermal) cutoff at $\sim 10-60\,$keV. The spectra also frequently display putative cyclotron absorption features (used to estimate the magnetic field strength), and a broadened iron emission line at $\sim 6\,$keV (e.g., White et al. 1983; Coburn et al. 2002).

Until recently, attempts to simulate the spectra of accretion-powered X-ray pulsars using first-principles physical models did not yield good agreement with the observed spectra. The earliest spectral models were based on the emission of blackbody radiation from the hot spots, and these were unable to reproduce the nonthermal power-law continuum. The presence of the cyclotron absorption feature led to the development of more sophisticated models based on a static slab geometry, in which the emitted spectrum is strongly influenced by cyclotron scattering (e.g., M\'esz\'aros \& Nagel 1985a,b; Nagel 1981; Yahel 1980b). While the magnetized slab models are able to roughly fit the shape of the observed cyclotron absorption features, they remained unable to reproduce the nonthermal power-law X-ray continuum.

The lack of a physics-based model for the formation of X-ray pulsar spectra led to the characterization of the observed spectra using parameters derived from a variety of {\it ad hoc} functional forms. Coburn et al. (2002) describe in detail three analytical functions commonly used to empirically model the X-ray continuum. The first function is the power law with high-energy cutoff (PLCUT), given by
\begin{equation}
    \textrm{PLCUT}(E) = AE^{-\Gamma}
    \left\{
    \begin{array}
    {l@{\quad \quad}l}
    1 \ , & (E \leq E_{\rm cut}) \ , \\
    e^{-(E-E_{\rm cut})/E_{\rm fold}} \ , & (E> E_{\rm cut}) \ ,
    \end{array}
    \right.
\label{eqn:PLCUT}
\end{equation}
where $E$ is the X-ray energy, $\Gamma$ is the photon spectral index, and $E_{\rm cut}$ and $E_{\rm fold}$ denote the cutoff and folding energies, respectively. The second function (Tanaka 1986) uses the same power law index $\Gamma$, combined with a Fermi-Dirac form for the high-energy cutoff (FDCO), defined by
\begin{equation}
\textrm{FDCO}(E) = AE^{-\Gamma}\frac{1}{1+e^{(E-E_{\rm cut})/E_{\rm fold}}} \ .
\label{eqn:FDCO}
\end{equation}
The third function (Mihara 1995) uses two power laws, $\Gamma_1$ and $\Gamma_2$, in combination with an exponential cutoff (NPEX), written as
\begin{equation}
\textrm{NPEX}(E) = (AE^{-\Gamma_1}+BE^{+\Gamma_2})e^{-E/E_{\rm fold}} \ .
\label{eqn:NPEX}
\end{equation}
Although these functional forms have no physical basis, they are often used to characterize the observed spectral shapes.

In later work, it eventually became clear that the observed power-law continuum was the result of a combination of bulk and thermal Comptonization occurring inside the accretion column. The first physically-motivated model based on these principles, that successfully described the shape of the X-ray continuum in accretion-powered pulsars, was developed by Becker \& Wolff (2007, hereafter BW07). This new model allowed for the first time the computation of the X-ray spectrum emitted through the walls of the accretion column based on the solution of a fundamental radiation-transport equation. The BW07 model was recently ported to {\it XSPEC} using a model described by Wolff et al. (2016). Another variation of the original BW07 model was developed and utilized by Farinelli et al. (2012; 2016, hereafter F16), which added an additional second-order scattering term to the transport equation related to the bulk flow. It is important to note that the Wolff et al. (2016) and Farinelli et al. (2012, 2016) implementations were premised on the utilization of an approximate form for the accretion velocity profile, which was assumed to stagnate at the stellar surface while varying in proportion to the scattering optical depth measured from the stellar surface. This approximate velocity profile stagnates at the stellar surface as required, but it does not follow the exact expected variation with altitude (Becker 1998).

While the BW07 and F16 models have demonstrated notable success in fitting the observed X-ray spectra for higher luminosity sources, these analytical models do not include all of the fundamental hydrodynamic and thermodynamic processes occurring within the accretion column. It is therefore natural to ask how the incorporation of additional physics would impact the ability of the BW07-type models to fit the observed X-ray spectra. This has motivated us to investigate the importance of additional radiative and hydrodynamical physics within the context of a detailed numerical simulation. The new simulation includes the implementation of a realistic dipole geometry, rigorous physical boundary conditions, and self-consistent energy transfer between electrons, ions, and radiation. In particular, we explore the dynamical effect of the gas pressure, which was neglected by BW07 and F16, in order to examine how the radiation and gas pressures combine to determine the dynamical structure of the accretion column and the associated radius-dependent and vertically-integrated, partially-occulted X-ray spectra.

This is the second in a series of two papers in which we describe the new coupled radiative-hydrodynamical model. The integrated approach involves an iteration between an ODE-based hydrodynamical code that determines the dynamical structure, and a PDE-based radiative code that computes the radiation spectrum. The iterative process converges rapidly to yield a self-consistent and unique description of the dynamical structure of the accretion column and the energy distribution in the emergent radiation field. In West et al. (2016, hereafter Paper I), we focus on solving the coupled hydrodynamical conservation equations to describe the underlying gas and radiation hydrodynamics within the accretion column. We employed {\it Mathematica} to solve the set of nonlinear hydrodynamical equations in dipole coordinates to determine the accreting bulk fluid velocity, the electron and ion temperature profiles, and the variation of the energy flux.

In this paper (Paper II) we focus on solving the fundamental PDE photon transport equation to describe the production of the observed radius-dependent and phase-averaged X-ray spectra. The bulk velocity and electron temperature profiles developed using the {\it Mathematica} calculation described in Paper I are used as input for the analysis conducted here, in which we solve the PDE photon transport equation using the {\it COMSOL} multiphysics module. The iterative dual-platform calculation is based on the transfer of information between the {\it COMSOL} radiation spectrum calculation and the {\it Mathematica} hydrodynamical structure calculation. Using this iterative approach, we are able to model realistic X-ray pulsar accretion columns in dipole geometry by solving the photon transport equation using self-consistent spatial distributions for the accretion velocity profile $v(r)$ and the electron temperature profile $T_e(r)$.

\newpage

\section{HYDRODYNAMICAL MODEL REVIEW}
\label{sec:HYDRODYNAMICAL MODEL REVIEW}

Our self-consistent model for the hydrodynamics and the radiative transfer occurring in X-ray pulsar accretion flows is based on a fundamental set of conservation equations governing the flow velocity, $v(r)$, the bulk fluid mass density, $\rho(r)$, the radiation energy density, $U_r(r)$, the ion energy density, $U_i(r)$, the electron energy density, $U_e(r)$, and the total radial energy transport rate, $\dot E(r)$, where $r$ is the radius measured from the center of the neutron star. In our hydrodynamical model, the field-aligned flow is decelerated by the combined pressure of the ions, the electrons, and the radiation. The total pressure is therefore given by the sum
\begin{equation}
P_{\rm tot} = P_e + P_i + P_r \ ,
\label{eqn:eos}
\end{equation}
where $P_e$, $P_i$ and $P_r$ denote the electron , ion, and radiation pressures, respectively. The equations-of-state for the material components are
\begin{equation}
P_i = n_i k T_i \ , \qquad
P_e = n_e k T_e \ .
\label{eqn:pressures}
\end{equation}
We note that in the X-ray pulsar application, the value of $P_r$ cannot be computed using a thermal relation because in general the radiation field is not in thermal equilibrium with either the matter or itself. Hence the radiation pressure must be computed using an appropriate conservation relation. The corresponding energy densities are related to the pressure components via
\begin{equation}
P_i = (\gamma_i-1) \, U_i \ , \quad P_e = (\gamma_e-1) \, U_e \ , \quad
P_r = (\gamma_r-1) \, U_r \ ,
\label{eqn:eos2}
\end{equation}
where (see Paper I) we set $\gamma_e=3$ and $\gamma_i=5/3$. The ratio of specific heats for the radiation field is given by $\gamma_r=4/3$.

As shown in Paper~I, the mathematical model can be reduced to a set of five coupled, first-order, nonlinear ordinary differential equations satisfied by $v$, $\dot E$, and the ion, electron, and radiation sound speeds, $a_i$, $a_e$, and $a_r$, respectively, defined by
\begin{equation}
a_i^2 = \frac{\gamma_i P_i}{\rho} \ , \quad
a_e^2 = \frac{\gamma_e P_e}{\rho} \ , \quad
a_r^2 = \frac{\gamma_r P_r}{\rho} \ .
\label{eqn:soundspeeds}
\end{equation}
The ion and electron temperatures, $T_i$ and $T_e$, respectively, are related to the corresponding sound speeds via
\begin{equation}
a_i^2 = \frac{\gamma_i k T_i}{m_{\rm tot}} \ , \ \ \ \
a_e^2 = \frac{\gamma_e k T_e}{m_{\rm tot}} \ ,
\label{eqn:temperatureconversion}
\end{equation}
where $m_i$ and $m_e$ denote the ion and electron masses, respectively, and $m_{\rm tot} \equiv m_e + m_i$. We assume here that the accreting gas is composed of pure, fully-ionized hydrogen.

\subsection{Conservation Equations}

Here, we briefly review the hydrodynamical conservation equations and the underlying model assumptions. The fundamental independent variables in our model are the photon energy, $\epsilon$, and the radius, $r$, measured from the center of the star. Since our model includes only one explicit spatial dimension, the structure across the column at a given value of $r$ is not treated in detail. Therefore, all of the physical quantities (velocity, temperature, pressure, flux, and density) developed here are interpreted as mean values across the column (see Appendix \ref{section:PHOTON ESCAPE FORMALISM}).

The mass continuity equation can be written in dipole geometry as (e.g., Langer \& Rappaport 1982)
\begin{equation}
\frac{\partial \rho(r)}{\partial t} = - \frac{1}{A(r)}\frac{\partial}{\partial r}
\Big[A(r) \rho(r) v(r) \Big] \ ,
\label{eqn:mass flux}
\end{equation}
where $v(r) < 0$ denotes the radial inflow velocity, and the cross-sectional area of the column, $A(r)$, is related to the solid angle $\Omega(r)$ subtended by the dipole column at radius $r$, by
\begin{equation}
A(r) = r^2 \Omega(r) = \frac{r^3 \, \Omega_*}{R_*} \ .
\label{eqn:area}
\end{equation}
The solid angle subtended by the column at the stellar surface is given by
\begin{equation}
\Omega_* = 2 \pi (\cos\theta_1 - \cos\theta_2) \ ,
\label{eqn:omega0}
\end{equation}
where $\theta_1$ and $\theta_2$ denote the inner and outer polar angles of the accretion column at the stellar surface, respectively. When $\theta_1=0$, the column is completely filled; otherwise, the central portion of the column is empty. We therefore find that the solid angle varies with radius according to
\begin{equation}
\Omega(r) = \left(\frac{r}{R_*}\right) \, \Omega_* \ .
\label{eqn:omega}
\end{equation}

In a steady-state situation, the mass accretion rate $\dot M$ is conserved, and is related to the density $\rho$ and velocity $v$ via
\begin{equation}
\dot M = \Omega \, r^2 \rho |v| \ .
\label{eqn:dipolarmassflux}
\end{equation}
Combining Equations~(\ref{eqn:omega}) and (\ref{eqn:dipolarmassflux}), we find that the variation of the mass density $\rho$ is given by
\begin{equation}
\rho(r) = \frac{\dot M R_*}{\Omega_* r^3 |v|} \ .
\label{eqn:massdensity}
\end{equation}
Under the assumption that the accreting gas is composed of pure, fully-ionized hydrogen, the electron and ion number densities are equal and can be computed using
\begin{equation}
n_e = n_i = \frac{\rho}{m_{\rm tot}} = \frac{\dot M R_*}{m_{\rm tot}
\Omega_* r^ 3 |v|} \ ,
\label{eqn:numberdensity}
\end{equation}
where $m_{\rm tot} = m_e + m_i$.

The total energy flux in the radial direction, averaged over the column cross-section at radius $r$, can be written as
\begin{equation}
F(r) = \frac{1}{2}\rho v^3 + v(P_i + P_e + P_r + U_i + U_e
+ U_r) - \frac{c}{n_e \sigma_\parallel}\frac{\partial P_r}{\partial r}
- \frac{GM_*\rho v}{r} \ ,
\label{eqn:energyflux}
\end{equation}
where $\sigma_\parallel$ denotes the electron-scattering cross section for photons propagating parallel to the magnetic field. The flux $F$ is defined to be negative for energy flow in the downward direction, towards the stellar surface, and the corresponding accretion velocity $v$ is negative ($v < 0$). The terms on the right-hand side of Equation~(\ref{eqn:energyflux}) represent the kinetic energy flux, the enthalpy flux, the radiation diffusion flux, and the gravitational energy flux, respectively. The diffusion of radiation energy through the walls is given by $d \dot E/d r$ and is modeled using an escape-probability formalism. The value of the energy transport rate $\dot E$ varies as a function of the radius $r$ in response to the escape of radiation perpendicular to the magnetic field direction. We therefore utilize a total energy conservation equation of the form
\begin{equation}
\frac{\partial}{\partial t}\left(\frac{1}{2}\,\rho v^2 + U_i + U_e + U_r
- \frac{G M_* \rho}{r} \right)
= - \frac{1}{A(r)}\frac{\partial\dot E}
{\partial r} - \frac{U_r}{t_{\rm esc}} \ ,
\label{eqn:energy1}
\end{equation}
where $\dot E$ is the total energy transport rate in the radial direction, defined by
\begin{equation}
\dot E \equiv A(r) F(r) \ ,
\label{eqn:Edot}
\end{equation}
and the mean escape timescale, $t_{\rm esc}$, is discussed in Section~\ref{sec:PHOTON TRANSPORT EQUATION} and in Appendix~\ref{section:PHOTON ESCAPE FORMALISM}.

It is convenient to work in terms of non-dimensional radius, flow velocity, sound speed, and total energy transport rate variables by introducing the quantities
\begin{equation}
\tilde r = \frac{r}{R_g} \ , \ \
\tilde v = \frac{v}{c} \ , \ \
\tilde a_i = \frac{a_i}{c} \ , \ \
\tilde a_e = \frac{a_e}{c} \ , \ \
\tilde a_r = \frac{a_r}{c} \ , \ \
\tilde{{\mathscr E}} = \frac{\dot E}{\dot M c^2} \ ,
\label{eqn:convertingr}
\end{equation}
where $R_g$ is the gravitational radius, defined by
\begin{equation}
R_g \equiv \frac{G M_*}{c^2} \ .
\label{eqn:RG}
\end{equation}
The computational domain extends from the top of the accretion column, at radius $\tilde r = \tilde r_{\rm top}$, down to the stellar surface at dimensionless radius $\tilde r=4.836$, which is based on canonical values for the stellar mass, $M_*=1.4\,M_\odot$, and the stellar radius, $R_*=10\,$km.

The set of five fundamental hydrodynamical differential equations that must be solved simultaneously using {\it Mathematica} comprises Equations~(\ref{eqn:RadiationODEtilde}), (\ref{eqn:IonODEtilde}), (\ref{eqn:ElectronODEtilde}), (\ref{eqn:EnergyODEtilde}), and (\ref{eqn:VelocityODEtilde}), which were described in detail in Paper I and are shown here as
\begin{align}
%% RadiationODEtilde
\frac{d \tilde a_r}{d \tilde r} &= \frac{\tilde a_r}{2} \left ( \frac{3}{\tilde r} + \frac{1}{\tilde v}\frac{d \tilde v}{d \tilde r} \right ) - \frac{\sigma_\parallel R_g}{2 m_{\rm tot} c} \frac{\dot M}{A} \frac{\gamma_r}{\tilde a_r} \left ( \tilde{\mathscr{E}} + \frac{\tilde v^2}{2} + \frac{\tilde a_e^2}{\gamma_e - 1} + \frac{\tilde a_i^2}{\gamma_i - 1} + \frac{\tilde a_r^2}{\gamma_r - 1} - \frac{1}{\tilde r} \right ), \label{eqn:RadiationODEtilde} \\
%% IonODEtilde
\frac{d \tilde a_i}{d \tilde r} &= \frac{(1 - \gamma_i) \tilde a_i}{2} \left ( \frac{3}{\tilde r} + \frac{1}{\tilde v}\frac{d \tilde v}{d \tilde r} + \frac{R_g}{c^2} \frac{A}{\dot M} \frac{\gamma_i \dot U_i}{\tilde a_i^2} \right ), \label{eqn:IonODEtilde} \\
%% ElectronODEtilde
\frac{d \tilde a_e}{d \tilde r} &= \frac{(1 - \gamma_e) \tilde a_e}{2} \left ( \frac{3}{\tilde r} + \frac{1}{\tilde v}\frac{d \tilde v}{d \tilde r} + \frac{R_g}{c^2} \frac{A}{\dot M} \frac{\gamma_e \dot U_e}{\tilde a_e^2} \right ), \label{eqn:ElectronODEtilde} \\
%% EnergyODEtilde
\frac{d \tilde{\mathscr{E}}}{d \tilde r} &= \frac{\tilde a_r^2}{c (\ell_2 - \ell_1) \gamma_r (\gamma_r - 1) \tilde v} \left ( \frac{R_*^3}{R_g \, \tilde r^3} \right)^{1/2} \min \left ( c, \frac{c}{\tau_\perp} \right) , \label{eqn:EnergyODEtilde} \\
%% VelocityODEtilde
\frac{d \tilde v}{d \tilde r} &= \frac{\tilde v}{\tilde v^2 -\tilde a_i^2 -\tilde a_e^2}
\left \{ \frac{3 \left ( \tilde a_i^2 +\tilde a_e^2 \right )}{\tilde r} - \frac{1}{\tilde r^2}
+\frac{\sigma_\parallel R_g}{m_{\rm tot} c} \frac{\dot M}{A}
\left ( \tilde{\mathscr{E}} + \frac{\tilde v^2}{2} + \frac{\tilde a_i^2}{\gamma_i-1}
+ \frac{\tilde a_e^2}{\gamma_e-1} \right . \right . \nonumber \\
& \qquad \qquad \qquad \qquad \left . \left . \ \ + \frac{\tilde a_r^2}{\gamma_r-1} - \frac{1}{\tilde r} \right )
+ \frac{R_g}{c^2} \frac{A}{\dot M}\left [ (\gamma_i - 1) \dot U_i + (\gamma_e - 1) \dot U_e \right ] \right \} \ , \label{eqn:VelocityODEtilde}
\end{align}
where the cross-sectional area $A(r)$ of the dipole accretion column is given by Equation~(\ref{eqn:area}). The ions do not radiate appreciably, and therefore the ion energy density is only affected by adiabatic compression and Coulomb energy exchange with the electrons (see Langer \& Rappaport 1982). Conversely, the electrons experience free-free, cyclotron, and Compton heating and cooling, in addition to adiabatic compression and the Coulomb exchange of energy with the ions. The radiation energy density responds to both the creation and absorption of photons via free-free and cyclotron processes, in addition to Compton scattering with the electrons.

The thermal coupling terms in Equations~(\ref{eqn:IonODEtilde}), (\ref{eqn:ElectronODEtilde}), and (\ref{eqn:VelocityODEtilde}) describe a comprehensive set of electron and ion heating and cooling processes which are broken down as follows,
\begin{eqnarray}
\dot U_e & = & \dot U_{\rm brem}^{\rm emit}
+ \dot U_{\rm brem}^{\rm abs}
+ \dot U_{\rm cyc}^{\rm emit}
+ \dot U_{\rm cyc}^{\rm abs}
+ \dot U_{\rm Comp} + \dot U_{\rm ei} \ , \nonumber \\
\dot U_i & = & - \dot U_{\rm ei} \ .
\label{eqn:Udottotal}
\end{eqnarray}
The terms in the expression for $\dot U_e$ denote, respectively, thermal bremsstrahlung (free-free) emission and absorption, cyclotron emission and absorption, the Compton exchange of energy between the electrons and photons, and electron-ion Coulomb energy exchange. In our sign convention, a heating term is positive and a cooling term is negative. The mathematical descriptions of each term in Equation~(\ref{eqn:Udottotal}) are provided in Paper~I. We note that the term $\dot U_{\rm Comp}$ plays a fundamental role in the energy exchange between photons and electrons, as discussed in detail below and in Section~\ref{sec:PHOTON TRANSPORT EQUATION}.

Thermal bremsstrahlung emission plays a significant role in cooling the ionized gas, and in the case of luminous X-ray pulsars, it also provides the majority of the seed photons that are subsequently Compton scattered to form the emergent X-ray spectrum. The electrons in the accretion column experience heating due to free-free absorption of low-frequency radiation, and they also experience heating and cooling due to the absorption and emission of thermal cyclotron radiation. These emission and absorption processes can play an important role in regulating the temperature of the gas. On average, however, cyclotron absorption does not result in net heating of the gas, due to the subsequently rapid radiative de-excitation, and we therefore set $\dot U_{\rm cyc}^{\rm abs}=0$ in our dynamical calculations. Near the surface of the accretion column, however, photons scattered out of the outwardly directed beam are not replaced, and this leads to the formation of the observed cyclotron absorption feature in a process that is very analogous to the formation of absorption lines in the solar spectrum (Ventura et al. 1979). While cyclotron absorption does not result in net heating of the gas, due to the rapid radiative de-excitation, cyclotron emission will cool the gas. The electrons can also be heated or cooled via Coulomb collisions with the protons, depending on whether the electron temperature $T_e$ exceeds the ion temperature $T_i$.

Compton scattering plays a fundamental role in the formation of the emergent X-ray spectrum. It is critically important in establishing the radial variation of the electron temperature profile through the exchange of energy between photons and electrons. The Compton cooling rate, prior to dimensionless units conversion, is expressed in terms of the mass density, $\rho$, the electron sound speed, $a_e$, and the radiation sound speed, $a_r$. The complete derivation of the electron cooling rate due to Compton scattering is given in Section~\ref{subsection:transport phenomena and photon-electron energetics}, and the final result is given by (see Equation~(\ref{eqn:Compton6b}))
\begin{equation}
\dot U_{\rm Comp}(r) = \frac{n_e \bar\sigma c}{m_e c^2} \ 4 k T_e \left[g(r) - 1 \right]
U_r(r) \ ,
\label{eqn:Udotcompton2a}
\end{equation}
where the temperature ratio function, $g(r)$, is defined by
\begin{equation}
g(r) \equiv \frac{T_{\rm IC}(r)}{T_e(r)} \ .
\label{eqn:gfunction}
\end{equation}
The derivation of $T_{\rm{IC}}$ is provided later in Section~\ref{subsection:transport phenomena and photon-electron energetics}. Equation~(\ref{eqn:Udotcompton2a}) can be stated in terms of the sound speeds as
\begin{equation}
\dot U_{\rm Comp} = \frac{4 \, \bar\sigma \, (g-1)}
{m_e c \gamma_e \gamma_r (\gamma_r-1)}
\ \rho^2 \, a_r^2 \, a_e^2 \ .
\label{eqn:Udotcompton2}
\end{equation}
The sign of $\dot U_{\rm Comp}$ depends on the value of $g$. The electrons experience Compton cooling if $g < 1$ (i.e. $T_{\rm IC} < T_e$); otherwise, the electrons are heated via inverse-Compton scattering.

Our task is to solve the five coupled hydrodynamic conservation equations, shown in Equations ({\ref{eqn:RadiationODEtilde})-(\ref{eqn:VelocityODEtilde}) above, to determine the radial profiles of the dynamic variables $\tilde a_r, \tilde a_i, \tilde a_e, \tilde{\mathscr{E}}$, and $\tilde v $, subject to the boundary conditions discussed in Section~\ref{subsection: Boundary Conditions} and Section~\ref{subsection: Transport Boundary Conditions}. Once these profiles are available, the electron temperature $T_e$ can be computed from the electron sound speed $\tilde a_e$ using the relation (see Equation (\ref{eqn:temperatureconversion}))
\begin{equation}
T_e = \frac{m_{\rm tot} c^2}{\gamma_e k} \, \tilde a_e^2 \ .
\label{eqn:Teequation}
\end{equation}
The solutions for $\tilde v(\tilde r)$ and $T_e(\tilde r)$ are the fundamental output of the solved system of conservation equations. These are used as input to the {\it COMSOL} finite-element environment in order to compute the photon distribution function, $f(r,\epsilon)$, inside the column, which is the focus of this paper.

Solving the five coupled conservation equations to determine the radial profiles of the quantities $v$, $\dot E$, $a_i$, $a_e$, and $a_r$ requires an iterative approach, because the rate of Compton energy exchange between the photons and the electrons depends on the relationship between the electron temperature, $T_e$, and the inverse-Compton temperature, $T_{\rm IC}$, which is determined by the shape of the radiation distribution (see Equation~(\ref{eqn:Compton7})). In order to achieve a self-consistent solution for all of the flow variables, while taking into account the feedback loop between the dynamical calculation and the radiative transfer calculation, the simulation must iterate through a specific sequence of steps. The steps required in a single iteration are: (1) the computation of the initial estimate of the accretion column dynamical structure, which is obtained by solving the five conservation equations described in Paper~I under the assumption that $T_{\rm IC}(r) = T_e(r)$; (2) calculation of the associated radiation distribution function by solving the radiative transfer equation described here in Paper~II; (3) computation of the inverse-Compton temperature profile $T_{\rm IC}(r)$ based on integration of the radiation energy distribution using Equation~(\ref{eqn:Compton7}); and (4) re-computation of the dynamical structure using the new estimate for $T_{\rm IC}(r)$. The iteration continues until adequate convergence is achieved between successive temperature profiles for all values of $r$. The iterative process is discussed in detail in Paper I, and also in Section~\ref{subsection:convergence} of this paper.

\subsection{Hydrodynamic Boundary Conditions}
\label{subsection: Boundary Conditions}

Here we summarize the set of boundary conditions utilized in the {\it Mathematica} integration of the set of hydrodynamical conservation equations, required in order to determine the structure of the accretion column. At the top of the accretion column (radius $r = r_{\rm top}$), the inflow velocity $v$ equals the local free-fall velocity, so that
\begin{equation}
v_{\rm top} \equiv v_{\rm ff}(r_{\rm top}) = -\left(\frac{2 G M_*}{r_{\rm top}}\right)^{1/2} \ .
\label{eqn:free-fall top}
\end{equation}
We also assume that the local acceleration of the gas is equal to the gravitational value, and therefore
\begin{equation}
\frac{dv}{dr}\bigg|_{r=r_{\rm top}} = \left(\frac{G M_*}{2 \, r_{\rm top}^3}\right)^{1/2} \ .
\label{eqn:free-fall acceleration top}
\end{equation}

At the upper surface of the dipole-shaped accretion funnel, the radiation escapes freely, forming a ``pencil-beam'' component in the observed radiation field. This physical principle can be used to derive a useful boundary condition at the upper surface of the accretion column. The radial component of the radiation energy flux, averaged over the cross-section of the column at radius $r$, is given in the diffusion approximation by (see Equation~(\ref{eqn:total radiation flux equation}))
\begin{equation}
F_r(r) = - \frac{c}{3 n_e \sigma_\parallel} \, \frac{dU_r}{dr} + \frac{4}{3}
\, v U_r \ ,
\label{eqn:radiation energy flux}
\end{equation}
where the first term on the right-hand side represents the upward diffusion of radiation energy parallel to the magnetic field, and the second term represents the downward advection of radiation energy towards the stellar surface (with $v < 0$). The fact that the top of the accretion column is the last scattering surface implies that the photon transport makes a transition from diffusion to free streaming at $r=r_{\rm top}$, and therefore we make the following replacement in Equation~(\ref{eqn:radiation energy flux}) at the upper boundary,
\begin{equation}
- \frac{c}{3 n_e \sigma_\parallel} \, \frac{dU_r}{dr} \to
c \, U_r \ , \qquad r \to r_{\rm top} \ .
\label{eqn:free-streaming}
\end{equation}
We incorporate Equation (\ref{eqn:free-streaming}) into Equation~(\ref{eqn:radiation energy flux}) to conclude that the radiation energy flux at the upper surface is given by
\begin{equation}
F_r(r) \bigg|_{r=r_{\rm top}} = \left ( c + \frac{4}{3} v_{\rm top} \right ) U_r(r_{\rm top}) \ ,
\label{eqn:radiation flux and free-streaming}
\end{equation}
where $v_{\rm top}$ is the free-fall velocity given by Equation~(\ref{eqn:free-fall top}).

The ionized gas flows downward after entering the top of the accretion funnel at radius $r=r_{\rm top}$, and eventually passes through a standing, radiation-dominated shock, where most of the kinetic energy is radiated away through the walls of the accretion column (Becker 1998). Below the shock, the gas passes through a sinking regime, where the remaining kinetic energy is radiated away (Basko \& Sunyaev 1976). Ultimately, the flow stagnates at the stellar surface, and the accreting matter merges with the stellar crust.

The surface of the neutron star is too dense for radiation to penetrate significantly (Lenzen \& Tr\"umper 1978), and therefore the sum of the diffusion and the advection components of the radiation energy flux must vanish there. Furthermore, due to the stagnation of the flow at the stellar surface, the advection component should also vanish, and we conclude the radiation energy flux $F_r \to 0$ as $r \to R_*$. We refer to this as the ``mirror"
surface boundary condition, which can be written as
\begin{equation}
F_r(r) \bigg|_{r=R_*} = 0 \ .
\label{eqn:radiation mirror condition}
\end{equation}
We approximate stagnation at the stellar surface in our simulations using the condition
\begin{equation}
\lim_{r \to R_*} |v(r)| \lapprox \ 0.01 \, c \ .
\label{eqn:surface stagnation}
\end{equation}

\section{PHOTON TRANSPORT EQUATION}
\label{sec:PHOTON TRANSPORT EQUATION}

The equation describing photon transport is a second order, non-linear, partial differential equation (PDE) of elliptic type, which can be solved numerically with appropriate boundary conditions using the finite-element method (FEM). The time-dependent transport equation for an isotropic photon distribution function $f$ (phase-space density of photons) experiencing Compton scattering, spatial diffusion, and bulk advection in an X-ray pulsar accretion column is given by the generalized Kompaneets (1957) equation (e.g., Becker 1992; Becker \& Wolff 2007)
\begin{eqnarray}
\frac{\partial f}{\partial t}
= \frac{n_e \bar\sigma c}{m_e c^2}\frac{1}{\epsilon^2}\frac{\partial}{\partial \epsilon}
\left[\epsilon^4 \left(f + k T_e \frac{\partial f}{\partial \epsilon}\right) \right]
- \vec\nabla \cdot \vec F_{\rm ph} + \dot f_{\rm prod} + \dot f_{\rm abs}
- \frac{1}{3 \, \epsilon^2}\frac{\partial}{\partial \epsilon}
\left(\epsilon^3 \,\vec v \cdot \vec\nabla f \right) \ ,
\label{eqn:fluxeqn1}
\end{eqnarray}
where the specific flux, or ``streaming,'' of the photons in the radial direction is given by (Gleeson \& Axford 1967; Skilling 1975)
\begin{equation}
\vec F_{\rm ph} = - \kappa \vec\nabla f - \frac{\epsilon \vec v}{3}\frac{\partial f}{\partial \epsilon} \ ,
\label{eqn:specificfluxvector}
\end{equation}
and $\kappa$ represents the spatial diffusion coefficient. The terms on the right-hand side of Equation~(\ref{eqn:fluxeqn1}) describe, respectively, the transport of the photons through the energy space due to stochastic electron scattering (the Kompaneets operator); the spatial diffusion of radiation; the injection of seed photons; the absorption of radiation by the accreting gas; and the differential work performed on the photons due to the convergence of the accreting gas (Gleeson \& Webb 1978; Cowsik \& Lee 1982). It is important to note that Equation~(\ref{eqn:fluxeqn1}) must be supplemented by suitable radiation boundary conditions for the photon distribution function $f$, imposed at the column walls, the stellar surface, and at the top of the column. This is further discussed in Section~\ref{subsection: Transport Boundary Conditions}.

Equations~(\ref{eqn:fluxeqn1}) and (\ref{eqn:specificfluxvector}) can be combined to obtain the single equivalent equation (e.g., Becker 2003)
\begin{eqnarray}
\frac{\partial f}{\partial t} + \vec v \cdot \vec\nabla f
= \frac{n_e \bar\sigma c}{m_e c^2}\frac{1}{\epsilon^2}\frac{\partial}{\partial \epsilon}
\left[\epsilon^4 \left(f + k T_e \frac{\partial f}{\partial \epsilon}\right) \right]
+ \vec\nabla \cdot \left(\kappa \vec\nabla f \right) \nonumber \\
+ \frac{1}{3} (\vec\nabla\cdot\vec v) \, \epsilon
\frac{\partial f}{\partial\epsilon} + \dot f_{\rm prod} + \dot f_{\rm abs} \ .
\label{eqn:PDEalt1}
\end{eqnarray}
In order to solve this equation, we must adopt a specific geometrical model. The accretion flow in an X-ray pulsar is channeled by the strong magnetic field, and therefore we will employ a dipolar geometry for the accretion column, in which the cross-sectional area of the column, $A(r)$, is given by Equation~(\ref{eqn:area}). The angle-dependent operators in Equation~(\ref{eqn:PDEalt1}) can be removed by averaging across the column cross-section, and by assuming azimuthal symmetry around the magnetic field axis. The resulting transport equation, satisfied by the isotropic photon distribution function $f(r,\epsilon)$ can be written in the dipole geometry as
\begin{eqnarray}
\frac{\partial f}{\partial t} + v \frac{\partial f}{\partial r}
= \frac{n_e \bar\sigma c}{m_e c^2}\frac{1}{\epsilon^2}\frac{\partial}{\partial \epsilon}
\left[\epsilon^4 \left(f + k T_e \frac{\partial f}{\partial \epsilon}\right) \right]
+ \frac{1}{A(r)}\frac{\partial}{\partial r}\left[A(r) \kappa \frac{\partial f}{\partial r} \right] \nonumber \\
+ \frac{1}{3 A(r)} \frac{\partial[A(r) v]}{\partial r} \, \epsilon \frac{\partial f}{\partial\epsilon}
+ \dot f_{\rm prod} + \dot f_{\rm abs} + \dot f_{\rm esc} \ ,
\label{eqn:PDEalt2}
\end{eqnarray}
where the spatial diffusion coefficient in the radial direction is given by
\begin{equation}
\kappa(r) = \frac{c}{3 n_e(r) \sigma_\parallel} \ .
\label{eqn:kappa}
\end{equation}

We demonstrate in Appendix~\ref{section:PHOTON ESCAPE FORMALISM} that the photon escape term $\dot f_{\rm esc}$ can be implemented using an escape probability formalism by writing
\begin{equation}
\dot f_{\rm esc}(r,\epsilon) = - \frac{f(r,\epsilon)}{t_{\rm esc}(r)} \ ,
\label{escape term in distribution function}
\end{equation}
where the mean escape time at radius $r$, denoted by $t_{\rm esc}(r)$, is given by
\begin{equation}
t_{\rm esc}(r) = \frac{\ell_{\rm esc}(r)}{w_\perp(r)} \ .
\label{eqn:escape time}
\end{equation}
Here, $\ell_{\rm esc}(r)$ denotes the perpendicular escape distance across the column at radius $r$, computed using
\begin{equation}
\ell_{\rm esc}(r) = (\ell_2 - \ell_1) \left(\frac{r}{R_*}\right)^{3/2} \ ,
\label{eqn:escape distance}
\end{equation}
and $w_\perp(r)$ is the perpendicular diffusion velocity, given by
\begin{equation}
w_\perp(r) = \min\left[c,\frac{c}{\tau_\perp(r)}\right] \ ,
\qquad
\tau_\perp(r) = n_e(r) \sigma_\perp \ell_{\rm esc}(r) \ ,
\label{eqn:diffusion velocity}
\end{equation}
where $\sigma_\perp$ denotes the electron-scattering cross section for photons propagating perpendicular to the magnetic field, and $\tau_\perp(r)$ is the corresponding perpendicular optical thickness of the accretion column at radius $r$. The parameters $\ell_1 = R_* \, \theta_1$ and $\ell_2 = R_* \, \theta_2$ in Equation~(\ref{eqn:escape distance}) represent the inner and outer arc-length surface radii at the base of the accretion column, respectively (see Equation~(\ref{eqn:omega0})). When $\ell_1=\theta_1=0$, the column is completely filled; otherwise, the center of the column is empty, which may reflect the manner in which the gas is entrained onto the field lines at large radii, or alternatively, it may reflect the possible presence of a quadrupole field component (see Section~\ref{sec:futurework}). Note that if $\tau_\perp < 1$, then $w_\perp(r)$ transitions to the free-streaming value, $w_\perp = c$, as required.

Significant variability of the X-ray emission on pulse-to-pulse timescales has been observed from some X-ray pulsars, including Her X-1 (Becker et al. 2012; Staubert et al. 2007; Staubert et al. 2014), in which the pulsation period is about one second. The pulsation period is several orders of magnitude larger than the dynamical timescale for accretion onto the star, and therefore, to first approximation, the accretion flow can be considered to be steady-state in the frame of the star. It is therefore sufficient for our purposes here to focus on solving the time-independent version of Equation~(\ref{eqn:PDEalt2}), although the steady-state assumption may be relaxed in future work. In order to reduce the complexity of the simulations while retaining the essential physical details, we will assume in this paper that all of the physical quantities represent averages across the column cross-section at a given value of the spherical radius $r$, measured from the center of the star. In future work, we plan to relax this assumption and treat the structure of the column using a more realistic two-dimensional model, but that is beyond the scope of the present paper. In the steady-state situation of interest here, Equation~(\ref{eqn:PDEalt2}) can be written in the flux-conservation form
\begin{eqnarray}
\frac{\partial}{\partial r}\left[\epsilon^2 r^3\left(-\kappa \frac{\partial f}{\partial r}
- \frac{v \epsilon}{3}\frac{\partial f}{\partial \epsilon} \right) \right]
+ \frac{\partial}{\partial \epsilon}
\left\{- \epsilon^2 r^3 \left[\frac{n_e \bar\sigma c}{m_e c^2} \,
\epsilon^2 \left(f + kT_e\frac{\partial f}{\partial \epsilon} \right)
- \frac{\epsilon v}{3}\frac{\partial f}{\partial r} \right] \right\} \nonumber \\
= \epsilon^2 r^3 \left(\dot f_{\rm prod} + \dot f_{\rm abs} - \frac{f}{t_{\rm esc}}\right) \ ,
\label{eqn:fluxvectoreqn}
\end{eqnarray}
where we have used Equations~(\ref{eqn:area}) and (\ref{escape term in distribution function}) to substitute for the cross-sectional area $A(r)$ and the escape term $\dot f_{\rm esc}$, respectively.

The solution space of the radiation transport equation contains spatial and energy dimensions, forming a 2D grid within the FEM-based {\it COMSOL} multiphysics environment. In the radial dimension, the computational domain extends from $r=R_*=10$\,km (the stellar surface) up to $r=r_{\rm top}\sim 20$\,km (the top of the accretion column). The precise value for $r_{\rm top}$ is determined self-consistently for each source, as discussed in Paper~I. In the energy dimension, the computational domain extends from a minimum photon energy $\epsilon_{\rm min} = 0.01$\,keV to a maximum photon energy $\epsilon_{\rm max} = 100$\,keV. The final form of the transport equation is found by converting Equation~(\ref{eqn:fluxvectoreqn}) to the dimensionless radius and velocity variables $\tilde r$ and $\tilde v$ (see Equation~(\ref{eqn:convertingr})). The result obtained is
\begin{eqnarray}
\frac{\partial}{\partial\tilde r} \left[\epsilon^2 R^2_g \tilde r^3\left(-\frac{c}{3 n_e \sigma_\parallel R_g}
\frac{\partial f}{\partial \tilde r}
- \frac{1}{3} c \tilde v \epsilon \frac{\partial f}{\partial \epsilon} \right)\right] \nonumber \\
+ \frac{\partial}{\partial \epsilon} \left\{R^3_g \tilde r^3 \epsilon^2 \left[- \frac{n_e \bar\sigma c}{m_e c^2}
\epsilon^2\left(f + kT_e \frac{\partial f}{\partial \epsilon}\right)
+ \epsilon \frac{c \tilde v}{3 R_g} \frac{\partial f}{\partial \tilde r} \right] \right\} \nonumber \\
= \epsilon^2 R^3_g \tilde r^3\left(\dot f_{\rm prod}
+ \dot f_{\rm abs} - \frac{f}{t_{\rm esc}} \right) \ ,
\label{eqn:finaltransportequation}
\end{eqnarray}
where we have also utilized Equation~(\ref{eqn:kappa}). Since {\it COMSOL} employs the finite-element method, it is well-suited to the iterative solution method required to solve the photon transport Equation~(\ref{eqn:finaltransportequation}), which is a second order, elliptic PDE. The associated boundary conditions are discussed in detail below, along with the forms utilized for the terms describing radiation absorption, injection, and escape.

\subsection{Photon Transport Boundary Conditions}
\label{subsection: Transport Boundary Conditions}

In order to solve the partial differential transport equation (Equation~(\ref{eqn:finaltransportequation})) for the photon distribution function $f$ in the {\it COMSOL} finite-element environment, we are obliged to develop and apply a suitable set of physical boundary conditions in radius and energy. The upper surface of the accretion column, located at radius $r = r_{\rm top}$, represents the last-scattering surface for photons diffusing up from deeper layers in the column. Hence at this radius, the photon transport makes a transition from classical diffusion to free-streaming, as occurs in the scattering layer above the photosphere in a stellar atmosphere. In the dipole geometry considered here, the specific flux or ``streaming'' in the radial direction is given by (cf. Equation~(\ref{eqn:specificfluxvector}))
\begin{equation}
F_{{\rm ph},\,r} \equiv - \kappa \frac{\partial f}{\partial r} - \frac{\epsilon v}{3}\frac{\partial f}{\partial \epsilon} \ .
\label{eqn:specificfluxvectorNEW}
\end{equation}
Using Equation~(\ref{eqn:kappa}) to substitute for the spatial diffusion coefficient $\kappa$, we find that the transition from diffusion to free-streaming implies that (cf. Equation~(\ref{eqn:free-streaming}))
\begin{equation}
- \frac{c}{3 n_e \sigma_\parallel} \, \frac{df}{dr} \to
c \, f \ , \qquad r \to r_{\rm top} \ ,
\label{eqn:free-streaming for f}
\end{equation}
and therefore
\begin{equation}
F_{{\rm ph},\,r} \to c \, f - \frac{\epsilon v}{3}\frac{\partial f}{\partial \epsilon} \ , \qquad r \to r_{\rm top} \ .
\label{eqn:freestreamtop}
\end{equation}
This relation expresses the spatial boundary condition applied to the photon distribution function $f$ at the column top.

We must also develop and apply an appropriate boundary condition at the stellar surface, where $r \to R_*$. In our idealized, one-dimensional model, the density of the accreting gas formally diverges as it settles onto the star, and therefore no radiation flux can penetrate the stellar surface. This concept is physically manifest in the ``mirror'' boundary condition, which states that the radiation streaming in the radial direction must vanish as $r \to R_*$. The mathematical implementation of this condition is given by
\begin{equation}
F_{{\rm ph},\,r} \to 0 \ , \qquad r \to R_* \ .
\label{eqn:mirrorbottom}
\end{equation}
The mirror condition described above provides the spatial boundary condition that must be satisfied by $f$ at the stellar surface.

The lowest and highest energies treated in our simulations are denoted by $\epsilon_{\rm min} = 0.01\,$keV and $\epsilon_{\rm max} = 100\,$keV, respectively. The boundary conditions applied at these two energies are determined by the requirement that the radiation transport rate along the energy axis must vanish in the limit $\epsilon \to \infty$, and also in the limit $\epsilon \to 0$. The numerical values we choose for $\epsilon_{\rm min}$ and $\epsilon_{\rm max}$ are such that we can assume that the radiation transport rate vanishes at these two energies, which implies that no photons cross the boundaries of the computational domain at $\epsilon = \epsilon_{\rm max}$ and $\epsilon = \epsilon_{\rm min}$. The high-energy boundary condition can be written as (cf. Equation~(\ref{eqn:fluxvectoreqn}))
\begin{eqnarray}
\frac{n_e \bar\sigma c}{m_e c^2} \,
\epsilon^2 \left(f + kT_e\frac{\partial f}{\partial \epsilon} \right)
- \frac{\epsilon v}{3}\frac{\partial f}{\partial r} = 0 \ , \qquad \epsilon = \epsilon_{\rm max} \ ,
\label{eqn:fluxvectoreqn2}
\end{eqnarray}
and likewise, the low-energy boundary condition is given by
\begin{eqnarray}
\frac{n_e \bar\sigma c}{m_e c^2} \,
\epsilon^2 \left(f + kT_e\frac{\partial f}{\partial \epsilon} \right)
- \frac{\epsilon v}{3}\frac{\partial f}{\partial r} = 0 \ , \qquad \epsilon = \epsilon_{\rm min} \ .
\label{eqn:fluxvectoreqn3}
\end{eqnarray}
Taken together, Equations~(\ref{eqn:freestreamtop}), (\ref{eqn:mirrorbottom}), (\ref{eqn:fluxvectoreqn2}), and (\ref{eqn:fluxvectoreqn3}) comprise the set of four boundary conditions that we apply on the four edges of the $(\epsilon,r)$ computational domain employed in the {\it COMSOL} environment.

\subsection{Photon Sources}
\label{subsection:photon sources}

The injection of seed photons due to the various emission mechanisms operating inside the accretion column is represented by the term $\dot f_{\rm prod}$ in Equation~(\ref{eqn:finaltransportequation}). Following BW07, we can express the photon production rate by writing
\begin{equation}
\dot f_{\rm prod} = \frac{Q_{\rm prod}}{r^2 \Omega(r)} = \frac{Q_{\rm brem} + Q_{\rm cyc}
+ Q_{\rm bb}}{r^2 \Omega(r)} \ ,
\label{eqn:fdot sources term}
\end{equation}
where the source functions $Q_{\rm brem}$, $Q_{\rm cyc}$, and $Q_{\rm bb}$ correspond to bremsstrahlung, cyclotron, and blackbody emission, respectively. The source functions are normalized so that $\epsilon^2 Q(r,\epsilon) \, d\epsilon \, dr$ gives the number of photons injected per unit time between radii $r$ and $r + dr$, with energy between $\epsilon$ and $\epsilon + d\epsilon$. The source function is therefore related to the volume emissivity, $\dot n_\epsilon$, via
\begin{equation}
\epsilon^2 Q(r,\epsilon) \, d\epsilon \, dr = r^2 \Omega(r) \, \dot n_\epsilon(r,\epsilon) \, d \epsilon \, dr \ ,
\label{eqn:newsourceemissivity}
\end{equation}
where $\dot n_\epsilon(r,\epsilon) \, d\epsilon$ gives the number of photons injected per unit time per unit volume in the energy range between $\epsilon$ and $\epsilon + d\epsilon$.

Computation of the thermal bremsstrahlung (free-free) source function, $Q_{\rm brem}$, is based upon Equation~(5.14b) from Rybicki \& Lightman (1979), which gives for the free-free volume emissivity (in cgs units)
\begin{equation}
\dot n_\epsilon^{\rm ff} = 3.7 \times 10^{36} \, \rho^2 T_e^{-1/2} \epsilon^{-1} e^{-\epsilon/kT_e} \ .
\label{eqn:bremsstrahlung photon emissivity}
\end{equation}
We can combine Equations~(\ref{eqn:newsourceemissivity}) and (\ref{eqn:bremsstrahlung photon emissivity}) to obtain for the bremsstrahlung source function
\begin{equation}
Q_{\rm brem} = 3.7 \times 10^{36} \, \Omega \, r^2 \rho^2 T_e^{-1/2} \epsilon^{-3} e^{-\epsilon/kT_e} \ .
\label{eqn:bremsstrahlung emission}
\end{equation}

The cyclotron volume emissivity, $\dot n_\epsilon^{\rm cyc}$, gives the production rate of cyclotron photons per unit volume per unit energy. For the case of pure, fully-ionized hydrogen, we can employ Equations~(7) and (11) from Arons et al. (1987) and substitute into Equation (\ref{eqn:newsourceemissivity}) to obtain
\begin{equation}
\dot n_{\epsilon}^{\rm cyc} = 2.1 \times 10^{36} \, \rho^2
B_{12}^{-3/2} H \left(\frac{\epsilon_{\rm cyc}}{k T_e}\right)
e^{-\epsilon_{\rm cyc}/k T_e} \delta \left(\epsilon-\epsilon_{\rm cyc} \right) \ ,
\label{eqn:cyclotron1}
\end{equation}
where $\epsilon_{\rm cyc}(r)=11.57\,B_{12}(r)\,$keV denotes the cyclotron energy, $B_{12}(r) \equiv B(r)/(10^{12}\,\rm G)$, and $H(x)$ is a piecewise function
defined by
\begin{equation}
H(x) \equiv
\begin{cases}
0.15 \, \sqrt{7.5} \ , & x \ge 7.5 \ , \\
0.15 \, \sqrt{x} \ , & x < 7.5 \ .
\end{cases}
\label{eqn:Hfunction}
\end{equation}
The radial variation of the magnetic field $B(r)$ is evaluated using Equation~(4) from Paper I, with $\theta = 0$, which gives
\begin{equation}
B(r) = B_* \left(\frac{r}{R_*}\right)^{-3} \ ,
\end{equation}
%.
where $B_*$ denotes the field strength at the center of the magnetic pole on the stellar surface. The cyclotron source function, $Q_{\rm cyc}$, is found by substituting Equation (\ref{eqn:cyclotron1}) into Equation (\ref{eqn:newsourceemissivity}), which gives
\begin{equation}
Q_{\rm cyc} = 2.1 \times10^{36} \, \Omega r^2 \rho^2 B_{\rm 12}^{-3/2} H \left( \frac{\epsilon_{\rm cyc}}{kT_e}\right)
\epsilon^{-2} e^{-\epsilon_{\rm cyc}/kT_e} \delta(\epsilon-\epsilon_{\rm cyc}) \ .
\label{eqn:cyclotron emission}
\end{equation}
In our numerical simulations, the $\delta$-function in Equation (\ref{eqn:cyclotron emission}) is approximated by a narrow Gaussian, centered on the cyclotron energy, with a standard deviation of 1 keV.

The blackbody source function, $Q_{\rm bb}$, is defined by writing
\begin{equation}
Q_{\rm bb} = S(\epsilon) \delta(r-r_{\rm th}) \ ,
\label{eqn:Qbb definition}
\end{equation}
where $r_{\rm th}$ is the radius at the upper surface of the thermal mound, and the function $S(\epsilon)$ is related to the Planck distribution according to
\begin{equation}
\epsilon^3 S(\epsilon) \, d\epsilon = r_{\rm th}^2 \Omega(r_{\rm th}) \, \pi B_\epsilon(\epsilon) \, d\epsilon \ ,
\label{eqn:relating S to blackbody intensity}
\end{equation}
with $B_\epsilon$ denoting the blackbody intensity, given by
\begin{equation}
B_\epsilon(\epsilon) = \frac{2 \epsilon^3}{c^2 h^3} \frac{1}{e^{\epsilon/k T_{\rm th}}-1} \ .
\label{eqn:Blackbody intensity}
\end{equation}
We can combine Equations~(\ref{eqn:Qbb definition}), (\ref{eqn:relating S to blackbody intensity}), and (\ref{eqn:Blackbody intensity}) to obtain for the blackbody photon source function the result
\begin{equation}
Q_{\rm bb}(r,\epsilon) = \frac{2\pi \, r_{\rm th}^2 \Omega(r_{\rm th})}{c^2 h^3} \frac{\delta(r-r_{\rm th})}
{e^{\epsilon/kT_{\rm th}}-1} \ .
\label{eqn:blackbodysource}
\end{equation}
The radial $\delta$-function in Equation~(\ref{eqn:blackbodysource}) is approximated using a narrow Gaussian feature, centered on the thermal mound, with a standard deviation of $0.207\,$km.

\subsection{Photon Absorption}
\label{subsection:photon production, escape, and absorption}

The photon absorption term, $\dot f_{\rm abs}$, in Equation~(\ref{eqn:finaltransportequation}) can be written in the form
\begin{equation}
\dot f_{\rm abs} = - \frac{f}{t_{\rm abs}} \ ,
\label{free-free absorption term in distribution function}
\end{equation}
where the mean absorption time, $t_{\rm abs}$, describes the average time for a photon to be reabsorbed through the bremsstrahlung absorption process (free-free absorption), given by
\begin{equation}
\frac{1}{t_{\rm abs}} = c \, \alpha_{\rm R} \ .
\label{eqn:absorption time}
\end{equation}
Here, $\alpha_{\rm R}$ denotes the Rosseland mean of the absorption coefficient for fully ionized hydrogen, expressed in cgs units by (Rybicki \& Lightman 1979)
\begin{equation}
\alpha_{\rm R} = 1.7 \times 10^{-25} \, T_e^{-7/2} n_e^2 \ \ \propto
\ \ {\rm cm}^{-1} \ .
\label{eqn:rosseland}
\end{equation}
The requirement of self-consistency in the treatment of the hydrodynamics and the radiative transfer is a major goal of the model developed here and in Paper~I, and therefore it is essential that we utilize the (frequency-independent) Rosseland mean absorption coefficient in Equation~(\ref{eqn:absorption time}).

\subsection{Transport Phenomena and Photon-Electron Energetics}
\label{subsection:transport phenomena and photon-electron energetics}

The angle-averaged (isotropic) photon distribution function, $f(r,\epsilon)$ is normalized so that $\epsilon^2 f(r,\epsilon) \, d\epsilon$ gives the number of photons per unit volume with energy between $\epsilon$ and $\epsilon+d\epsilon$ inside the accretion column at radius $r$, measured from the center of the star. From knowledge of $f$, we can compute the radiation number density $n_r$ and energy density $U_r$ using the relations
\begin{equation}
n_r(r) = \int_0^\infty \epsilon^2 f(r,\epsilon) \, d \epsilon \ , \ \ \ \ \
U_r(r) = \int_0^\infty \epsilon^3 f(r,\epsilon) \, d \epsilon \ .
\label{eqn:Compton5}
\end{equation}
We derive the ordinary differential equations satisfied by $n_r$ and $U_r$ in Appendix \ref{section:PHOTON ENERGY DENSITY ODE} by integrating the full transport equation using Equations~(\ref{eqn:Compton5}). The radiation energy flux, $F_r$ (erg cm$^{-2}$ s$^{-1}$), is defined by
\begin{equation}
F_r(r) = -\frac{c}{3 n_e \sigma_\parallel} \frac{dU_r}{dr} + \frac{4}{3} v U_r \ ,
\label{eqn:total radiation flux equation}
\end{equation}
which is obtained by integrating the radial component of the specific flux, $F_{\rm ph}$, via (see Equation~(\ref{eqn:specificfluxvector}))
\begin{equation}
F_r(r) = \int_0^\infty \epsilon^3 F_{\rm ph}(r,\epsilon) \, d \epsilon \ .
\label{eqn:RadEnergyFlux}
\end{equation}

Compton scattering plays a dominant role in determining the nature of the thermal balance between the electrons and the radiation field. In a single scattering, the mean value of the energy transferred from the photon to the electron is given by (Rybicki \& Lightman 1979)
\begin{equation}
\left<\Delta\epsilon\right> = \frac{\epsilon}{m_e c^2} (4 k T_e - \epsilon) \ ,
\label{eqn:Compton8}
\end{equation}
where $\epsilon$ is the photon energy. The $\dot U_{\rm Comp}$ term in Equation~(\ref{eqn:Udottotal}) is the rate of change of the electron energy density $U_e$ due to Compton scattering. Hence this quantity can be computed by writing
\begin{equation}
\dot U_{\rm Comp}(r) = - n_e \bar\sigma c \int_0^\infty \epsilon^2 f(r,\epsilon)
\, \left<\Delta\epsilon\right> \, d\epsilon \ ,
\label{eqn:Compton4}
\end{equation}
where the negative sign appears because the electrons {\it gain} energy whenever $\left<\Delta\epsilon\right> < 0$ for the photons. Combining Equations~(\ref{eqn:Compton8}) and (\ref{eqn:Compton4}) yields
\begin{equation}
\dot U_{\rm Comp}(r) = \frac{n_e \bar\sigma c}{m_e c^2} \left[
\int_0^\infty \epsilon^4 f(r,\epsilon)
\, d\epsilon - 4 k T_e \int_0^\infty
\epsilon^3 f(r,\epsilon) \, d\epsilon \right] \ ,
\label{eqn:Compton6}
\end{equation}
which can be written in the equivalent form
\begin{equation}
\dot U_{\rm Comp}(r) = \frac{n_e \bar\sigma c}{m_e c^2} \ 4 k \left[T_{\rm IC}(r) - T_e(r)\right]
U_r(r) \ ,
\label{eqn:Compton6b}
\end{equation}
where we have used Equation~(\ref{eqn:Compton5}) to introduce $U_r$, and we have also made use of the definition of the inverse-Compton temperature, $T_{\rm IC}$, given by (Rybicki \& Lightman 1979)
\begin{equation}
T_{\rm IC}(r) \equiv \frac{1}{4k}\frac{\int_0^\infty
\epsilon^4 f(r,\epsilon) \, d\epsilon}{\int_0^\infty \epsilon^3 f(r,\epsilon)
\, d\epsilon} \ .
\label{eqn:Compton7}
\end{equation}
Note that the electrons lose energy to the photons when $T_e > T_{\rm IC}$, and they gain energy from the photons when $T_e < T_{\rm IC}$. In the special case $T_e = T_{\rm IC}$, there is no net exchange of energy between the photons and electrons via Compton scattering.

The fundamental reason for the iteration between {\it Mathematica} and {\it COMSOL} is that the pressure distribution in the column, which determines the dynamics, is strongly influenced by the Compton exchange of energy between the radiation and the plasma, which depends on the radial profile of the inverse-Compton temperature $T_{\rm IC}(r)$ relative to the electron temperature $T_e(r)$. Since the radiation field is generally far from equilibrium in X-ray pulsar accretion columns, it follows that $T_{\rm IC}$ is not necessarily equal to $T_e$, which we demonstrate here in Paper~II.

From an operational point of view, the inverse-Compton temperature profile $T_{\rm IC}(r)$ must be computed in {\it COMSOL} from the radiation energy distribution $f(r,\epsilon)$ inside the accretion column by applying Equation~(\ref{eqn:Compton7}). This vector of information is output from {\it COMSOL}, and input into the {\it Mathematica} simulation of the hydrodynamical structure, as discussed in detail in Paper~I. The {\it Mathematica} integration results in the next iteration for the structure of the accretion column, which includes the velocity profile $\tilde v(r)$, the electron temperature profile $T_e(r)$, and the electron, ion, and radiation sound speed profiles, denoted by $\tilde a_e(r)$, $\tilde a_i(r)$, and $\tilde a_r(r)$, respectively. These dynamical profiles are then output from {\it Mathematica} and input into {\it COMSOL} in order to compute the next iteration of the radiation distribution function $f(r,\epsilon)$. This iterative process is repeated until the inverse-Compton temperature profile converges, as discussed in detail below.

\section{COMPUTING THE PHOTON SPECTRUM}
\label{sec:COMPUTING THE PHOTON SPECTRUM}

The solution of the set of hydrodynamic ODEs and the associated photon transport equation requires the specification of six fundamental free parameters, namely the angle-averaged electron scattering cross section, $\bar\sigma$, the parallel electron scattering cross section, $\sigma_\parallel$, the inner and outer polar cap radii, $\ell_1$ and $\ell_2$, the incident radiation Mach number, ${\mathscr M}_{r0}$, and the surface field strength at the magnetic pole, $B_*$. The values of the six free parameters are varied in order to obtain the best (visual) match of the simulated (computed) spectrum, when compared with the observed phase-averaged spectrum for a given source. The results obtained for the six fundamental free parameters are listed in Table~\ref{tab:free parameters for model sources} for each of the three sources treated here.

There are an additional thirteen parameters, either constrained or derived, which complete the total set of nineteen model parameters that uniquely describe a solution. The six constrained parameters are based upon the inherent properties of the pulsar itself, and comprise the distance to the source $D$, the observed luminosity $L_{\rm X}$, the scattering cross section for photons propagating perpendicular to the field $\sigma_\perp$, the accretion rate $\dot M$, and the stellar mass $M_*$ and radius $R_*$. We assume canonical values for the stellar mass and radius, given by $R_* = 10$ km and $M_* = 1.4\,M_\odot$, respectively. The distance to the source, $D$, is an observational parameter that is usually determined via association of the source with a globular cluster of known distance, or by direct measurement using very-long baseline interferometry. The values obtained for the six constrained parameters are reported in Table~\ref{tab:constrained parameters for model sources}, and in Table~\ref{tab:Auxiliary Parameters} we document the values used for the various auxiliary parameters associated with the modeling of the iron emission, cyclotron absorption, and blackbody emission from the accretion disk. The scattering cross section for photons propagating perpendicular to the magnetic field is dominated by the ordinary mode (Arons et al. 1987), and therefore we set the perpendicular scattering cross section equal to the Thomson value, $\sigma_\perp = \sigma_{\rm T}$.

The approximate X-ray luminosity, $L_{\rm X}$, is taken from published estimates, and the corresponding mass accretion rate, $\dot M$, is computed by setting $L_{\rm X}$ equal to the accretion luminosity, $L_{\rm acc}$, defined by
\begin{equation}
L_{\rm acc} \equiv \frac{G M_* \dot M}{R_*} \ ,
\end{equation}
which assumes that the heating of the star during the accretion process is negligible. Finally, the remaining seven parameters, listed in Table~6 from Paper~I, are derived quantities that are functions of the six free parameters and the boundary conditions, as discussed in Section~\ref{subsection: Boundary Conditions}.

The resulting photon distribution function $f(r,\epsilon)$ is a central result of this paper. All transport phenomena are calculated based on $f(r,\epsilon)$, including the total radiation energy flux inside the column, $F_r(r)$, the radiation energy density, $U_r(r)$, and the photon number density, $n_r(r)$. In this paper we present both phase-averaged and radius-dependent X-ray spectra computed using our model for Her X-1, Cen X-3, and LMC X-4. We compare the resulting X-ray spectra with the observational data for these three sources in order to exercise the model using sources spanning a wide range of luminosities, and to gain new insight into the emission phenomena occurring in X-ray pulsar accretion columns.

\subsection{Distribution Function Convergence}
\label{subsection:convergence}

Our method for determining the convergence of the numerical solution is based on the comparison of successive iterates of the temperature profiles $T_e(r)$ and $T_{\rm IC}(r)$. We define the convergence ratios, ${\mathscr R}_e(r)$ and ${\mathscr R}_{\rm IC}(r)$, respectively, for the electron and inverse-Compton temperatures using
\begin{equation}
{\mathscr R}_e^{n+1}(r) \equiv \frac{T_e^{n+1}(r)}{T_e^n(r)} \ , \qquad
{\mathscr R}_{\rm IC}^{n+1}(r) \equiv \frac{T_{\rm IC}^{n+1}(r)}{T_{\rm IC}^n(r)} \ ,
\label{eqn:convergence}
\end{equation}
where the superscripts represent the iteration number for the corresponding solution vectors. The solutions are deemed to have converged when the radial vector of convergence ratios for both the electron and the inverse-Compton temperature profiles are within 1\% of unity across the entire computational grid. The spatial grid spans the distance from the top of the accretion column, at $r = r_{\rm top}$ (or dimensionless radius $\tilde r = \tilde r_{\rm top}$), to the stellar surface at $r = R_*$ (or dimensionless radius $\tilde r = 4.836$, assuming canonical neutron star parameters). Once convergence is achieved, we have obtained a self-consistent set of results for the radiation distribution function $f(r,\epsilon)$, and also for the five dynamical variables $\tilde v(r)$, $\tilde a_r(r)$, $\tilde a_i(r)$, $\tilde a_e(r)$, and $\tilde{\mathscr E}(r)$.

\subsection{Interstellar and Cyclotron Absorption}
\label{subsection:cyclotron and interstellar absorption}

Soft X-ray absorption by the interstellar medium was first discussed by Bell \& Kingston (1967), Brown \& Gould (1970), and Charles et al. (1973). In this process, the X-ray intensity from the source is attenuated according to
\begin{equation}
I(\epsilon) = I_0 \, A_{\rm NH}(\epsilon) \ ,
\qquad
A_{\rm NH}(\epsilon) \equiv e^{-{\rm N_H} \sigma_n(\epsilon)} \ ,
\label{eqn:attenuation1}
\end{equation}
where ${\rm N_H}$ ($\propto$ cm$^{-2}$) denotes the intervening column density of neutral hydrogen atoms, and $\sigma_n(\epsilon)$ is the net photoelectric absorption cross-section per H nucleus. Figure~1 in Morrison \& McCammon (1983) provides the net photoelectric absorption cross-section per hydrogen atom as a function of energy for typical elemental abundances in the interstellar medium. The analytic fit to the net cross-section is approximated using
\begin{equation}
\sigma_n(\epsilon) = 10^{-24} {\rm cm}^2 \times \left[\frac{c_0(\epsilon_{\rm keV})
+ c_1(\epsilon_{\rm keV}) \, \epsilon_{\rm keV}
+ c_2(\epsilon_{\rm keV}) \, \epsilon_{\rm keV}^2}{\epsilon_{\rm keV}^3} \right] \ ,
\label{eqn:photo-electric cross-section per H nucleus}
\end{equation}
where $\epsilon_{\rm keV}$ denotes the photon energy in keV units, and the coefficients $c_0(\epsilon)$, $c_1(\epsilon)$, and $c_2(\epsilon)$ in Equation (\ref{eqn:photo-electric cross-section per H nucleus}) have constant values within each of the 14 energy bins (spanning the range from 0.03\,keV - 10\,keV) indicated in Table~2 from Morrison \& McCammon (1983). Outside the energy range 0.03\,keV - 10\,keV, the photoelectric cross-section $\sigma_n = 0$, and $A_{\rm NH}(\epsilon)=1$.

The cyclotron resonant scattering feature is modeled as a 2D Gaussian function (e.g., Heindl \& Chakrabarty 1999; Orlandini et al. 1998; Soong et al. 1990), with both spatial and energy dependences, and is superimposed directly onto the spectrum by multiplying by the attenuation function, $A_{\rm cyc}(r,\epsilon)$, defined using
\begin{equation}
A_{\rm cyc}(r,\epsilon) \equiv 1 - \frac{d_r}{\sigma_r \sqrt{2 \pi}} e^{-(r - r_{\rm cyc})^2/(2 \sigma_r^2)} \frac{d_c}{\sigma_{\rm cyc} \sqrt{2 \pi}} e^{-(\epsilon - \epsilon_{\rm cyc})^2/(2 \sigma_{\rm cyc}^2)} \ ,
\label{eqn:Ac Gaussian}
\end{equation}
where $r_{\rm cyc}$ is the cyclotron absorption imprint radius, $\epsilon_{\rm cyc}$ is the cyclotron absorption centroid energy, $d_r$ and $d_c$ are strength parameters, and $\sigma_{\rm cyc}$ and $\sigma_r$ are standard deviations, respectively. We can combine the two strength parameters into a single joint strength parameter, defined as $d_{cr} = d_c \, d_r$, which reduces Equation (\ref{eqn:Ac Gaussian}) to the form
\begin{equation}
A_{\rm cyc}(r,\epsilon) = 1-\frac{d_{cr}}{2 \pi \sigma_{\rm cyc} \sigma_r} e^{-(\epsilon - \epsilon_{\rm cyc})^2
/(2 \sigma_{\rm cyc}^2)} e^{-(r - r_{\rm cyc})^2/(2 \sigma_r^2)} \ .
\label{eqn:2D Gaussian final form}
\end{equation}
The specific values used for the strength parameter, $d_{cr}$, and the standard deviations, $\sigma_{\rm cyc}$ and $\sigma_r$, for each source are provided in Section~\ref{section:ASTROPHYSICAL APPLICATIONS}, and in Section~\ref{section:DISCUSSION AND CONCLUSION} we discuss the validity of using such an approximation.

\subsection{Fan-Beam and Pencil-Beam Components}
\label{subsection:altitude-dependent spectra}

The photon transport equation is solved in {\it COMSOL} using a high-density finite-element 2D meshed grid, comprising one radial dimension and one energy dimension. This yields the numerical solution for the fundamental photon distribution function, $f(r,\epsilon)$, from which we can calculate the inverse-Compton temperature profile $T_{\rm IC}(r)$, the radiation flux $F_r(r)$, the photon number density $n_r(r)$, and the energy density $U_r(r)$. Once the iterative process is completed and convergence is achieved, we can calculate the spectrum emerging through the column walls in two ways. The first option is to obtain an approximation of the phase-averaged photon count rate spectrum, $F_\epsilon(\epsilon)$, by integrating the ``fan beam'' component of the escaping spectrum with respect to radius $r$ over the entire length of the accretion column, from $r=r_{\rm top}$ to $r=R_*$. The second option is to explore the effects of partial occultation of the column due to the stellar rotation, as seen by a distant observer, which is accomplished by integrating the fan component from the top of the column down to a selected radius, $r$, which is above the stellar surface, so that $r > R_*$. In addition to the fan component, we can also compute the ``pencil beam'' component that emanates from the upper surface of the accretion column, at radius $r = r_{\rm top}$.

Bremsstrahlung, cyclotron, and blackbody emission each make contributions to the escaping radiation spectrum, as discussed in Section~\ref{subsection:photon sources}. These emission components are linked due to the inherent nonlinearity of the radiation transport equation (Equation~(\ref{eqn:finaltransportequation})). The nonlinearity reflects the fact that the electron temperature profile, $T_e(r)$, is influenced by the inverse-Compton temperature of the radiation field, $T_{\rm IC}(r)$, which in turn depends on the shape of the radiation spectrum via Equation~(\ref{eqn:Compton7}). Despite the interconnection between the Comptonized bremsstrahlung, cyclotron, and blackbody components, once the iterative process has converged, and the electron temperature profile $T_e(r)$ is known, we can then calculate the separate contributions to the observed spectrum due to each seed photon source. This comparison is carried out in Section~\ref{section:ASTROPHYSICAL APPLICATIONS}.

Following BW07, we can express the contribution to the fan-beam component of the photon number spectrum resulting from the escape of photons through the sides of the column between radii $r$ and $r + dr$ using
\begin{equation}
\dot N_\epsilon(r, \epsilon) =
\frac{\Omega(r) \, r^2 \epsilon^2}{t_{\rm esc}(r)} f(r,\epsilon)
\quad \propto \quad {\rm s^{-1} \, cm^{-1} \, keV^{-1}} \ ,
\label{eqn:Ndot}
\end{equation}
where the mean escape timescale, $t_{\rm esc}(r)$, is computed using Equation (\ref{eqn:escape time}). Note that the quantity $\dot N_\epsilon(r, \epsilon) \, d\epsilon\, dr$ represents the number of photons escaping per unit time through the column side walls between radii $r$ and $r+dr$ in the energy range between $\epsilon$ and $\epsilon+d\epsilon$.

Once the attenuation factors $A_{\rm NH}(\epsilon)$ and $A_{\rm cyc}(r,\epsilon)$ have been computed using Equations~(\ref{eqn:attenuation1}) and (\ref{eqn:2D Gaussian final form}), respectively, we can calculate the fan-beam component of the observed photon number flux spectrum measured in the reference frame of a distant observer, $F_\epsilon(\epsilon)$, due to radiation escaping from a sub-domain of the column between radii $r$ and $r_{\rm top}$ using the integral
\begin{equation}
F_\epsilon(\epsilon) = \frac{A_{\rm NH}(\epsilon)}{4 \pi D^2} \int_{r}^{r_{\rm top}}
A_{\rm cyc}(r,\epsilon) \, \dot N_\epsilon (r,\epsilon) \, dr
\quad \propto \quad {\rm s^{-1} \, cm^{-2} \, keV^{-1}} \ ,
\label{eqn:wall phase-averaged photon count rate spectrum}
\end{equation}
where $D$ is the distance to the source. The sub-domain integrated spectrum given by Equation~(\ref{eqn:wall phase-averaged photon count rate spectrum}) can used to compute the phase-averaged and the partially-occulted fan-beam radiation components detected by a distant observer. For example, the result for $F_\epsilon(\epsilon)$ corresponds to the approximate phase-averaged spectrum if we integrate $\dot N_\epsilon$ over the entire column length, from $r = r_{\rm top}$ to $r = R_*$. Alternatively, we can integrate $\dot N_\epsilon$ over a smaller radial sub-domain that extends from the top of the column ($r = r_{\rm top}$) down to any selected radius $r$. In this case, the result for $F_\epsilon(\epsilon)$ approximates the spectrum detected by a distant observer when the column is partially occulted by the star's surface due to the stellar rotation.

The individual, radius-dependent fan-beam components due to the escape of Comptonized bremsstrahlung, cyclotron, and blackbody emission are denoted by $\dot N^{\rm brem}_\epsilon(r,\epsilon)$, $\dot N^{\rm cyc}_\epsilon(r,\epsilon)$, and $\dot N^{\rm bb}_\epsilon(r,\epsilon)$, respectively. Once these functions have been generated, the total (pre-absorption) fan-beam photon number spectrum is then computed using the sum
\begin{equation}
\dot N^{\rm tot}_\epsilon(r,\epsilon) = \dot N^{\rm brem}_\epsilon(r,\epsilon)
+ \dot N^{\rm cyc}_\epsilon(r,\epsilon) + \dot N^{\rm bb}_\epsilon(r,\epsilon)
\quad \propto \quad {\rm s^{-1} \, cm^{-1} \, keV^{-1}} \ ,
\label{eqn:Ndot components}
\end{equation}
The corresponding fan-beam components of the observed photon spectrum, including absorption, are given by
\begin{equation}
F_\epsilon^{\rm tot}(\epsilon) = F_\epsilon^{\rm brem}(\epsilon) + F_\epsilon^{\rm cyc}(\epsilon)
+ F_\epsilon^{\rm bb}(\epsilon)
\quad \propto \quad {\rm s^{-1} \, cm^{-2} \, keV^{-1}} \ ,
\label{eqn:phase-averaged photon count rate spectrum}
\end{equation}
where each term on the right-hand side is computed by replacing $\dot N_\epsilon(r,\epsilon)$ in Equation~(\ref{eqn:wall phase-averaged photon count rate spectrum}) with the appropriate solution for each photon source, and then carrying out the integration with respect to radius. Depending on the integration bounds used in Equation~(\ref{eqn:wall phase-averaged photon count rate spectrum}), the sum over emission components in Equation (\ref{eqn:phase-averaged photon count rate spectrum}) can provide the numerical prediction for either the phase-averaged or the partially-occulted X-ray flux spectrum observed from an accretion column. We will compare the predicted phase-averaged spectrum with previously published data for three luminous sources in Section~\ref{section:ASTROPHYSICAL APPLICATIONS}. We will also investigate the contributions due to each seed photon source by examining the individual components in Equation~(\ref{eqn:phase-averaged photon count rate spectrum}).

The observed spectrum will be the sum of the fan-beam component (Equation~(\ref{eqn:phase-averaged photon count rate spectrum})) and the pencil-beam component, created by the escape of radiation through the free-streaming surface at the top of the column. The pencil-beam emission from the column top, $\hat F_\epsilon(\epsilon)$, including absorption, is computed using
\begin{equation}
\hat F_\epsilon(\epsilon) = \frac{A_{\rm NH}(\epsilon) \Omega(r_{\rm top})
\, r_{\rm top}^2 c \, \epsilon^2}{4 \pi D^2} f(r_{\rm top},\epsilon)
\quad \propto \quad {\rm s^{-1} \, cm^{-2} \, keV^{-1}} \ .
\label{eqn:column top photon count rate spectrum}
\end{equation}
In the case of the pencil-beam component, only interstellar absorption is included, because cyclotron absorption is concentrated at radius $r_{\rm cyc}$, which is about half the radius at the column top, $r_{\rm top}$ (see Tables~6 and 7 from Paper~I). The total pencil-beam emission spectrum, representing the escape of Comptonized photons from the top of the column at radius $r = r_{\rm top}$ due to all three emission mechanisms, is given by the sum (cf. Equation~(\ref{eqn:phase-averaged photon count rate spectrum}))
\begin{equation}
\hat F_\epsilon^{\rm tot}(\epsilon) = \hat F_\epsilon^{\rm brem}(\epsilon) + \hat F_\epsilon^{\rm cyc}(\epsilon)
+ \hat F_\epsilon^{\rm bb}(\epsilon)
\quad \propto \quad {\rm s^{-1} \, cm^{-2} \, keV^{-1}} \ ,
\label{eqn:top photon count rate spectrum}
\end{equation}
with each component on the right-hand side computed by substituting the corresponding solution for $f$ into Equation~(\ref{eqn:column top photon count rate spectrum}).

\section{ASTROPHYSICAL APPLICATIONS AND DISCUSSION}
\label{section:ASTROPHYSICAL APPLICATIONS}

In this section, we use the new model to study the three X-ray pulsars Her X-1, Cen X-3, and LMC X-4, which have relatively high luminosities in the range $L_{\rm X} \sim 10^{36-38}\,\rm erg\,s^{-1}$. The values obtained for the six model free parameters, $\bar\sigma$, $\sigma_\parallel$, $\ell_1$, $\ell_2$, ${\mathscr M}_{r0}$, and $B_*$, are listed in Table~\ref{tab:free parameters for model sources}, and the values for the constrained parameters are listed in Table~\ref{tab:constrained parameters for model sources}. We compare the theoretical predictions for the emergent column-integrated spectra with the observed phase-averaged spectral data for each of the three pulsars in the photon energy range from 0.1\,keV to 100\,keV.

There are a variety of emission components that add to create the total phase-averaged X-ray spectrum. Our model provides separate output components for the fan-beam and pencil-beam spectral components of bremsstrahlung, cyclotron, and blackbody seed photons, while additional visible spectral features due to iron emission and cyclotron resonant absorption are added using Gaussian approximations. In the case of Her X-1, there is also a soft component included in the model, which is interpreted as emission from the accretion disk. The visual matching procedure for extracting model parameters and solving for the phase-averaged photon count rate spectrum, $F_\epsilon(\epsilon)$, is divided into four basic steps:

\begin{deluxetable}{llll}
\tablewidth{0pt}
\tablecaption{Free Parameters for Her X-1, Cen X-3, and LMC X-4
\label{tab:free parameters for model sources}}
\tablehead{\colhead{Parameter} & \colhead{Her X-1} & \colhead{Cen X-3} &
\colhead{LMC X-4}}
\startdata
Angle-averaged cross section $\bar{\sigma}/\sigma_{\rm T}$ & $2.60\times10^{-3}$ & $3.00 \times 10^{-3}$ & $2.50\times10^{-3}$ \\
Parallel scattering cross section $\sigma_\parallel/\sigma_{\rm T}$ & $1.02\times10^{-3}$ & $7.51 \times 10^{-4}$ & $4.18\times10^{-4}$ \\
Inner polar cap radius $\ell_1$ (m) & 0 & 657 & 547 \\
Outer polar cap radius $\ell_2$ (m) & 125 & 750 & 650 \\
Incident radiation Mach $\Mach_{r0}$ & 4.07 & 6.15 & 2.76 \\
Surface magnetic field $B_*$ ($10^{12}\,$G) & 6.25 & 3.60 & 8.00 \\
\enddata
\end{deluxetable}

\begin{deluxetable}{lllll}
\tablewidth{0pt}
\tablecaption{Constrained Parameters for Her X-1, Cen X-3, and LMC X-4
\label{tab:constrained parameters for model sources}}
\tablehead{\colhead{Parameter} & \colhead{Her X-1} & \colhead{Cen X-3} & \colhead{LMC X-4} & \colhead{units}}
\startdata
$R_*$ & $10$ & $10$ & $10$ & km\\
$M_*$ & $1.4 \, M_\odot$ & $1.4 \, M_\odot$ & $1.4 \, M_\odot$ & g\\
$D$ & 5.0 & 8.0 & 55.0 & kpc\\
$L_{\rm X}$ & $2.00\times10^{37}$ & $2.82\times10^{38}$ & $3.89\times10^{38}$ & erg s$^{-1}$\\
$\dot M$ & $1.08 \times 10^{17}$ & $1.52 \times 10^{18}$ & $2.09 \times 10^{18}$ & g s$^{-1}$\\
$\sigma_\perp$ & $\sigma_{\rm T}$ & $\sigma_{\rm T}$ & $\sigma_{\rm T}$ & cm$^2$\\
\enddata
\end{deluxetable}

$\bullet$ Step 1: Once a source is selected, provisional values are assigned to the six fundamental free parameters, $\bar\sigma$, $\sigma_\parallel$, $\ell_1$, $\ell_2$, ${\mathscr M}_{r0}$, and $B_*$. The remaining 13 constrained and derived parameters are then computed.

$\bullet$ Step 2: {\it Mathematica} is used to solve the coupled hydrodynamical ODE system, comprising Equations (\ref{eqn:RadiationODEtilde}) through (\ref{eqn:VelocityODEtilde}), obeying all boundary conditions. We initially assume (the ``0$^{\rm th}$ iteration") that the electron temperature, $T_e(r)$, and the inverse-Compton temperature, $T_{\rm IC}(r)$, are equal for all values of $r$, so that $\dot U_{\rm Comp} = 0$ (see Equation~(\ref{eqn:Compton6b})). This is an input assumption for the first {\it Mathematica} run, which results in an output electron temperature vector $T_e(r)$ along with an output velocity vector $v(r)$. These two vectors are then exported from {\it Mathematica} into {\it COMSOL} in preparation for the ``0$^{\rm th}$ iteration" solution of the photon transport PDE, discussed below.

$\bullet$ Step 3: {\it COMSOL} is used to solve the photon transport equation (Equation (\ref{eqn:finaltransportequation})) to obtain the photon distribution function $f(r,\epsilon)$ (photons cm$^{-3}$ keV$^{-3}$). The inverse-Compton temperature profile $T_{\rm IC}(r)$ is then calculated using Equation (\ref{eqn:Compton7}). At this point, it becomes clear that $T_e(r)$ and $T_{\rm IC}(r)$ are not necessarily equal, and therefore the energy transfer via the $\dot U_{\rm Comp}$ term can become substantial. The radius-dependent fan-beam component, $\dot N_\epsilon(r,\epsilon)$, is computed using Equation~(\ref{eqn:Ndot}), and the corresponding phase-averaged fan-beam count rate spectrum $F_\epsilon(\epsilon)$ (photons s$^{-1}$ cm$^{-2}$ keV$^{-1}$) is calculated using Equation (\ref{eqn:wall phase-averaged photon count rate spectrum}). We also compute the pencil-beam component, $\hat F_\epsilon(\epsilon)$, using Equation~(\ref{eqn:column top photon count rate spectrum}).

$\bullet$ Step 4: The values of the convergence ratios ${\mathscr R}_e(r)$ and ${\mathscr R}_{\rm IC}(r)$ defined in Equations (\ref{eqn:convergence}) are computed. The model is considered unconverged if either $|1-{\mathscr R_e}| > 0.01$ or $|1-{\mathscr R}_{\rm IC}| > 0.01$ for at least one value of the grid radius $r$. In this case, the updated $T_{\rm IC}(r)$ vector is returned to {\it Mathematica} for inclusion in the coupled ODE system, and the iteration procedure loops back to Step 1. The model has converged when the ratios of the most recently calculated $T_{\rm IC}(r)$ and $T_e(r)$ profiles, compared with their previous respective profiles, differ by less than 1\% across the entire radial grid. In this case, the calculation terminates, and the resulting phase-averaged spectrum is compared with the observational data for the source in question. If necessary, the free parameters listed in Table~\ref{tab:free parameters for model sources} are varied, and the iterative procedure is restarted at Step 1.

\subsection{Her X-1}

Our primary findings regarding the dynamical and radiative properties of Her X-1 are depicted in Figure \ref{fig:Her X-1 phase-averaged properties}. The bulk fluid velocity ($\tilde v_{\rm bulk}$) and plasma temperature profiles ($T_i$ and $T_e$) are discussed in greater detail in Paper I. The fluid enters the top of the column in a free-fall condition at an altitude of 11.19\,km above the stellar surface, satisfying momentum conservation with a free-fall velocity gradient. The radiation sound speed ($\tilde a_r$) eventually rises above the decelerating bulk fluid at the radiation sonic surface, which in the case of Her X-1 is at a distance of $\sim$2\,km above the stellar surface. Below the radiation sonic surface the bulk fluid rapidly decelerates into an extended sinking regime (Basko \& Sunyaev 1976). Eventually the bulk fluid approaches the stellar surface with a residual stagnation velocity $v_* = -0.0084\,c$.

The ions and electrons are nearly in equilibrium with each other along the entire column, which is consistent with the short timescale for ion and electron equilibration (see also Paper I). Above the radiation sonic surface, the photons are continuously adding heat to the electrons via Compton scattering (i.e., $T_{\rm IC} > T_e$), and we observe that the average photon energy slowly decreases as the flow descends over a distance of $\sim 8$\,km. Below the sonic surface, the roles are reversed, and the highly compressed bulk fluid transfers energy from electrons to the photons via inverse-Compton scattering ($T_{\rm IC} < T_e$). In response, the mean photon energy increases as the fluid enters the sinking regime. The radiation energy flux, $F_r$, is a combination of downward advection and upward diffusion (see Equation~\ref{eqn:total radiation flux equation}), and is essentially equal to zero at the stellar surface, in satisfaction of the ``mirror'' boundary condition discussed in Section~\ref{subsection: Boundary Conditions}. In the case of Her X-1, the imprint radius of the cyclotron resonance scattering feature, denoted by $r_{\rm cyc}$, is equal to $r_{\rm X}$, which is defined as the radius at which the emergent luminosity per unit length, $\mathscr{L}_r$, is maximized (see Equation~(120) from Paper I).

We can gain a deeper understanding of the energy exchange between the photons and electrons by investigating the radial variation of the Compton $y$-parameter (Rybicki \& Lightman 1979), which describes whether an average photon experiences a significant change in energy as it scatters through the column before escaping. In general, a $y$-parameter exceeding unity implies that saturated Comptonization is occurring, leading to the appearance of a distinctive Wien bump in the spectrum. The thermal Compton $y$-parameter is calculated using
\begin{equation}
y_{\rm thermal} = \frac{t_{\rm esc}}{\epsilon} \left< \frac{d\epsilon}{dt} \right>_{\rm thermal} \ ,
\label{eqn:thermal $y$-parameter}
\end{equation}
where repeated Compton scattering of photons with energy $\epsilon \ll kT_e$ results in photon energization with the mean rate
\begin{equation}
\left< \frac{d\epsilon}{dt} \right>_{\rm thermal} = n_e \bar \sigma c \, \epsilon \, \frac{4 k T_e}{m_e c^2} \ .
\label{eqn:photon thermal energization rate}
\end{equation}
Bulk fluid compression also results in photon energization, with a mean rate given by
\begin{equation}\label{eqn:photon bulk energization rate}
\left< \frac{d\epsilon}{dt} \right>_{\rm bulk} = - \left(\vec\nabla \cdot \vec v \right) \frac{\epsilon}{3} \ ,
\end{equation}
and therefore the bulk $y$-parameter is found using
\begin{equation}
y_{\rm bulk} = \frac{t_{\rm esc}}{\epsilon} \left< \frac{d\epsilon}{dt} \right>_{\rm bulk} \ .
\label{eqn:bulk $y$-parameter}
\end{equation}
In Her X-1, bulk Comptonization is the dominant mode of photon energization above the radiation sonic surface, whereas thermal Comptonization begins to dominate just below the sonic surface, at an altitude of $\sim 1.5$\,km from the stellar surface. Both $y$-parameters are less than unity until the fluid approaches the sonic surface, which establishes that the defining characteristics of the Her X- spectrum are developed within the last 2\,km of the accretion flow, in a region where where thermal Comptonization dominates over bulk compression.

\begin{figure}[htbp]
\centering
\includegraphics[width=6.0in]{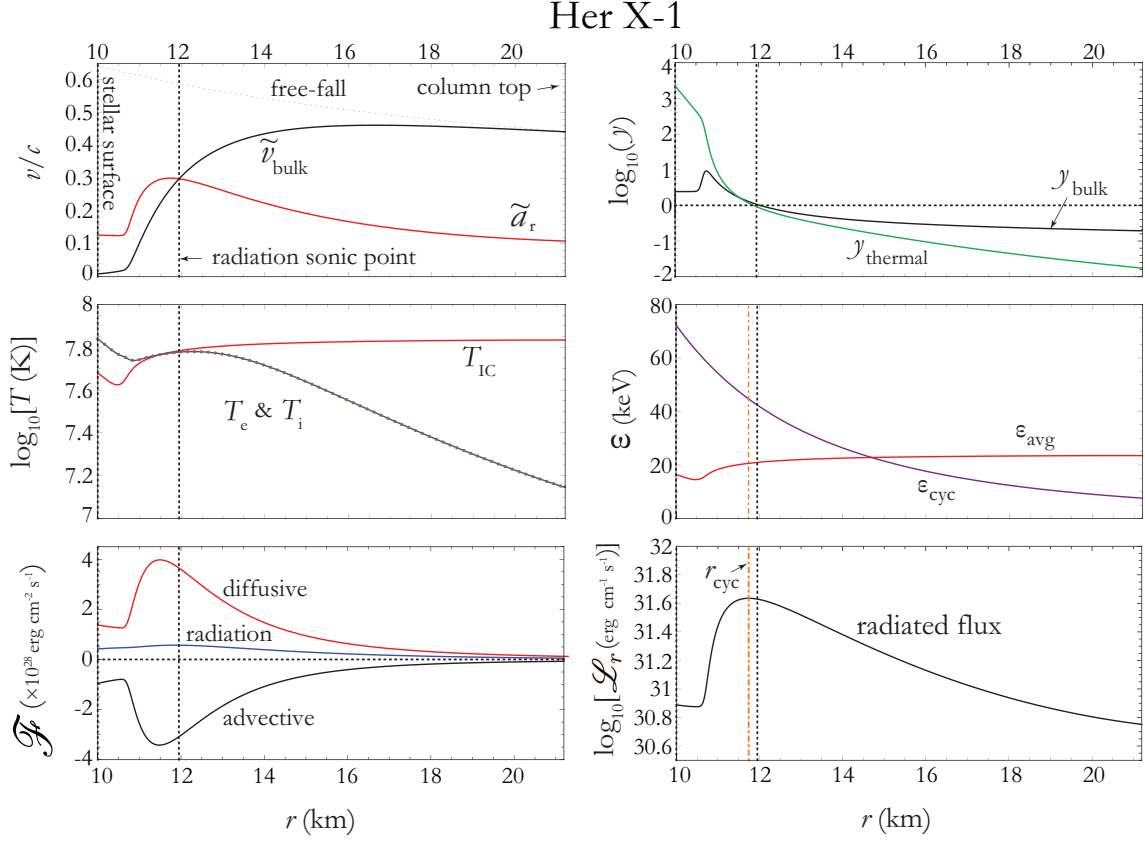}
\caption[Her X-1 properties]{Phase-averaged dynamical and radiative properties for Her X-1. The locations of the thermal mound, the cyclotron imprint radius, and the radiation sonic point are indicated. See the discussion in the text.}
\label{fig:Her X-1 phase-averaged properties}
\end{figure}

Figure \ref{fig:Her X-1 radius-dependent and partially-occulted spectra} depicts the radius-dependent (lower panel) and partially-occulted (upper panel) spectra for Her X-1. The radius-dependent spectra correspond to the spectra emitted per cm at a precise value of the radius $r$, whereas in the partially-occulted case, the value of $r$ corresponds to the lower bound of integration in Equation~(\ref{eqn:wall phase-averaged photon count rate spectrum}). We use four values for the radius in Figure \ref{fig:Her X-1 radius-dependent and partially-occulted spectra}; the thermal mound surface (green), radiation sonic surface (blue), and at distances from the stellar surface equal to 1/2 and 3/4 of the column length (red and orange), respectively. The lower panel depicts the fan-beam component only (the top spectrum is not included), radiated per cm of the column length. In addition to a contribution from the pencil-beam at the top of the column, each spectrum plotted in the upper panel includes two iron emission lines and an additional blackbody component that dominates at low energies. These additional spectral contributions are discussed further below. For illustration purposes, interstellar absorption and cyclotron absorption are included only in the vertically-integrated, partially-occulted spectrum in the upper panel.

The radius-dependent contributions are very uniform at energies below $\sim 10\,$keV. The maximum emitted flux, at all energies, occurs in the vicinity of the radiation sonic location at 1.95\,km above the stellar surface. The power-law shape of the spectrum is dominated by emission escaping from the lowest regions of the column, whereas emission from the upper half of the column has essentially no impact on the observed spectral shape. The photon dynamics change above 10\,keV, however, and all of the spectra above the radiation sonic surface show a slightly positive bump in the spectral magnitude, whereas there is a clear suppression in high energy photons at the thermal mound. The radial integration to obtain the partially-occulted spectra in the upper panel of Figure~\ref{fig:Her X-1 radius-dependent and partially-occulted spectra} begin with the pencil beam emission component as the starting point, and increase as the integration continues downwards towards the stellar surface. The partially-occulted spectra begin to exhibit the effects from cyclotron absorption at an altitude of 5.6\,km above the stellar surface, which is about half of the column length in this case.

The phase-averaged photon count rate spectrum for Her X-1 is calculated using Equation (\ref{eqn:phase-averaged photon count rate spectrum}) and plotted in Figure \ref{fig:Her X-1 phase-averaged spectrum} over the energy range 0.1-100\,keV. We note that the phase-averaged spectrum is equivalent to the partially-occulted spectrum, if the lower integration bound $r$ is set equal to the stellar radius $R_*$ in Equation~(\ref{eqn:wall phase-averaged photon count rate spectrum}). The upper panel in Figure \ref{fig:Her X-1 phase-averaged spectrum} depicts the total spectrum without the effects of interstellar absorption and cyclotron absorption included. The lower panel includes both absorption effects. The red dots represent the incident phase-averaged {\it BeppoSAX} data reported by Dal Fiume et al. (1998) with the instrumental response removed. The cumulative contribution to the total spectrum is shown as the solid black curve that includes blackbody (brown), cyclotron (blue), and bremsstrahlung (green) emission from the side walls (solid) and the top (dashed) of the column.

The spectra plotted in Figure~\ref{fig:Her X-1 phase-averaged spectrum} include two iron emission lines, which are approximated using Gaussian functions centered at energies $\epsilon_{\rm K1} = 6.45\,$keV and $\epsilon_{\rm K2} = 0.96\,$keV (Oosterbroek et al. 1997, Dal Fiume et al. 1998). The auxiliary parameters listed in Table~\ref{tab:Auxiliary Parameters} document the values used for the iron emission line centroid energies, as well as the corresponding standard deviations ($\sigma_{\rm K1}$ and $\sigma_{\rm K2}$), and the line strength parameters ($d_{\rm K1}$ and $d_{\rm K2}$). A 2D Gaussian cyclotron absorption feature (see Equation (\ref{eqn:2D Gaussian final form})) is also included, centered at an imprint radius $r_{\rm cyc} = 11.74$\,km, corresponding to a cyclotron centroid energy $\epsilon_{\rm cyc} = 44.72\,$keV. Implementation of the 2D Gaussian cyclotron absorption feature requires the specification of the spatial standard deviation, $\sigma_r=9.2$\,km, the energy standard deviation, $\sigma_c=1$\,keV, and the strength parameter, $d_{cr}=353$ (see Table~\ref{tab:Auxiliary Parameters}). A discussion of these values, and the general manner in which absorption is treated in our model, is provided in Section~\ref{section:DISCUSSION AND CONCLUSION}. Interstellar absorption best matches the observed Her X-1 data when we adopt for the column density $N_{\rm H} = 5.25\times10^{19} {\rm cm}^{-2}$, which agrees closely with the spectral fit parameter $N_{\rm H} = 5.10 \pm 0.7 \, (\times10^{19} {\rm cm}^{-2})$ obtained by Dal Fiume et al. (1998). The parameters for the 2D cyclotron Gaussian for all sources are listed in Table~\ref{tab:Auxiliary Parameters}, in addition to the parameters describing the hydrogen column density, the iron emission, and, in the case of Her X-1, the surface area and temperature of the excess soft blackbody component, which we discuss below.

\begin{deluxetable}{lccc}
\tablewidth{0pt}
\tablecaption{Auxiliary Parameters
\label{tab:Auxiliary Parameters}}
\tablehead{\colhead{Parameter} & \colhead{Her X-1} & \colhead{Cen X-3} & \colhead{LMC X-4}}
\startdata
Log$_{10}(N_{\rm H})$ (cm$^{-2}$) & 19.72  & 22.20  & 21.97\\
$\epsilon_{\rm cyc}$ (keV)                & 44.72  & 31.79  & 32.15\\
$\sigma_{\rm cyc}$ (keV)                  & 11.10  & 11.50  & 11.30\\
$\sigma_r$ (km)                   & 9.20   & 5.17   & 4.14\\
$d_{cr}$                          & 353    & 216    & 120\\
$\epsilon_{\rm K1}$ (keV)         & 6.45   & 6.67   & 5.90\\
$\sigma_{\rm K1}$ (keV)           & 0.400  & 0.293  & 0.190\\
$d_{\rm K1}$                      & 0.0060 & 0.0084 & 0.0007\\
$\epsilon_{\rm K2}$ (keV)         & 0.96   & N/A    & N/A\\
$\sigma_{\rm K2}$ (keV)           & 0.157  & N/A    & N/A\\
$d_{\rm K2}$                      & 0.028  & N/A    & N/A\\
$\sigma_{\rm bb}$ (km)            & 0.207  & 0.207  & 0.207\\
Blackbody area (cm$^2$)           & $9 \times 10^{15}$ & N/A & N/A\\
Blackbody temperature (Kelvin)    & $1.06 \times 10^6$ & N/A & N/A\\
\enddata
\end{deluxetable}

As early as 1975, it was well-documented that Her X-1 has an intense soft X-ray flux at energies below $\sim 1\,$keV (Shulman et al. 1975; Catura \& Acton 1975). We also find clear evidence for this excess, which cannot be explained as direct emission from the accretion column. In order to visually match the observed phase-averaged spectrum, we include a blackbody component with temperature $T = 1.06 \times 10^6{\rm K}$ ($kT = 0.091\,$keV) and radiating surface area $9.0 \times 10^{15}\,{\rm cm}^2$. These values agree with the results of Catura \& Acton (1975), who identified a blackbody component with temperature $T = 10^6$\,K and radiating surface area $9.85 \times 10^{15}\,{\rm cm}^2$. Note that this area is about three orders of magnitude larger than the entire surface area of the neutron star, and therefore it probably originates in the accretion disk, or, alternatively, it may represent emission due to Compton scattering of the primary X-rays by the coronal gas above and below the accretion disk (Jones \& Forman 1976; Becker et al. 1977, Pravdo et al. 1977, Bai 1980). For example, Bai (1980) proposed that a clumpy distribution of coronal gas above and below the accretion disk at a distance $\sim 10^8$\,cm could explain the soft X-ray contribution, which agrees with our average Alfv\'en radius of $\sim 5.3\times10^8$\,cm (see Paper I). Oosterbroek et al. (1997) suggested that the blackbody component may be reprocessed emission originating from regions of large optical depth within the accretion disk. Endo et al. (2000) obtained a spectral fit containing a 0.16\,keV blackbody component with an emitting surface radius of 240\,km, at a distance similar to the expected inner radius of the accretion disk. Ramsay et al. (2002) calculated a thermal blackbody component at $\sim1$keV with a similar large emitting radius located between $\sim 140$\,km and 500\,km in the accretion disk. Ji et al. (2009) calculated a 0.114\,keV thermal blackbody source with an emitting radius between 10\,km to 100\,km, and concluded that the soft component likely comes from the inner edge of the accretion disk. Finally, F\"urst et al. (2013) calculated a radius of $\sim$250\,km for an emitting blackbody at 0.14\,keV. These results are consistent with the properties of the soft blackbody component included in our model for Her X-1.

The X-ray emission components depicted in Figure~\ref{fig:Her X-1 phase-averaged spectrum} include all of the individual pencil-beam and fan-beam contributions due to the Comptonization of bremsstrahlung, cyclotron, and blackbody seed photons. The phase-averaged spectrum is clearly dominated by radiation from the wall (fan beam), while the low-energy spectrum is dominated by the blackbody component discussed in the previous paragraph. A number of authors have concluded that Her X-1 is being viewed from a direction roughly perpendicular to the magnetic field (e.g., M\'esz\'aros 1977). Simplistically the emission is ``cup-shaped" (Bisnovatyi-Kogan et al. 1976). A new and significant prediction from our model is that the pencil-beam component is comparable to the fan-beam component close to the cyclotron absorption feature.

\begin{figure}[htbp]
\centering
\includegraphics[width=5.0in]{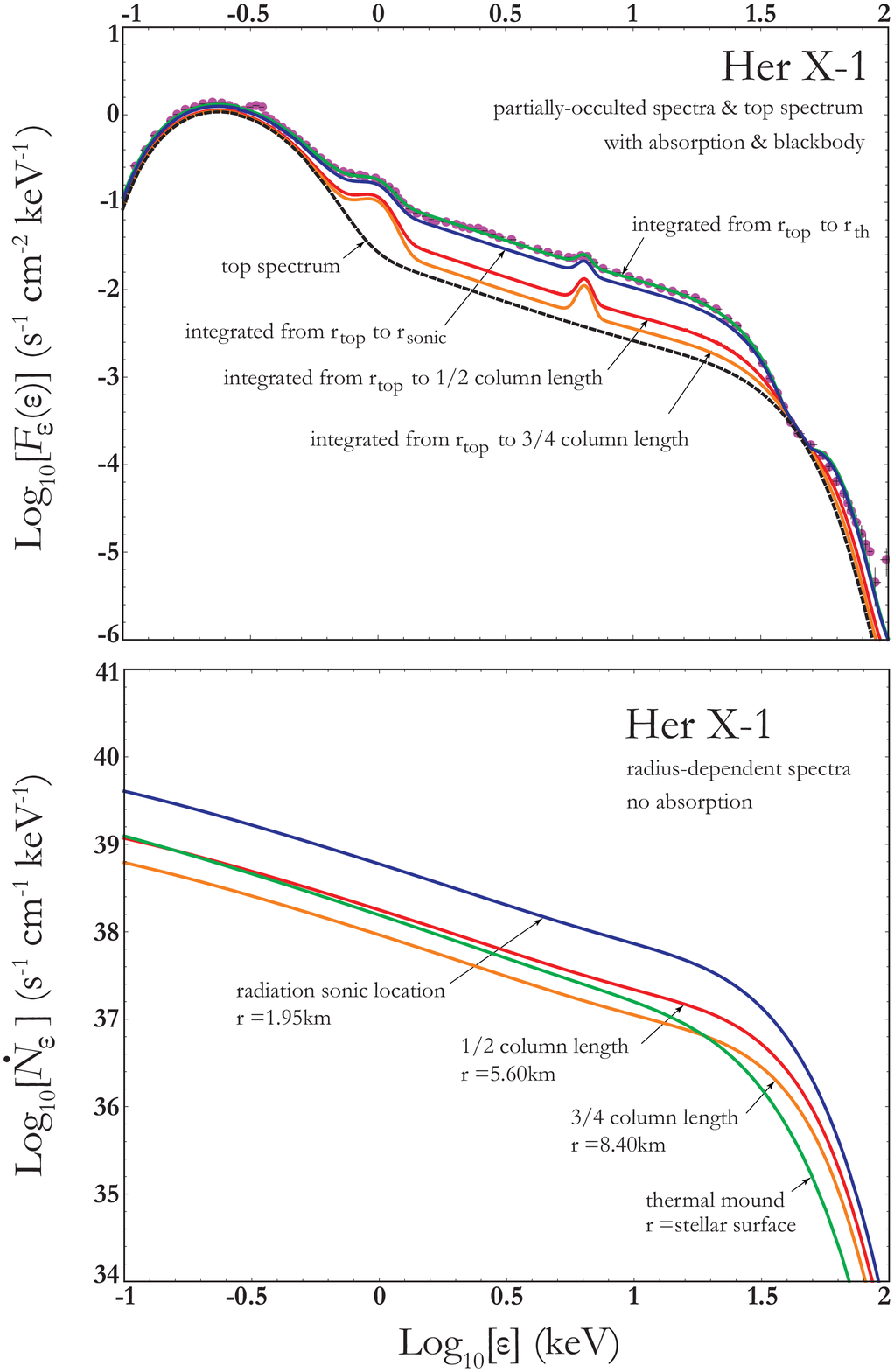}
\caption[Her X-1 Spectrum]{Partially-occulted spectra computed using Equation (\ref{eqn:wall phase-averaged photon count rate spectrum}), integrated over the indicated range of radius $r$ (upper panel). Radius-dependent escaping number spectrum at the indicated value of $r$, computed using Equation (\ref{eqn:Ndot}) (lower panel). The upper panel also includes the pencil-beam component computed using Equation~(\ref{eqn:column top photon count rate spectrum}).}
\label{fig:Her X-1 radius-dependent and partially-occulted spectra}
\end{figure}
\begin{figure}[htbp]
\centering
\includegraphics[width=5.0in]{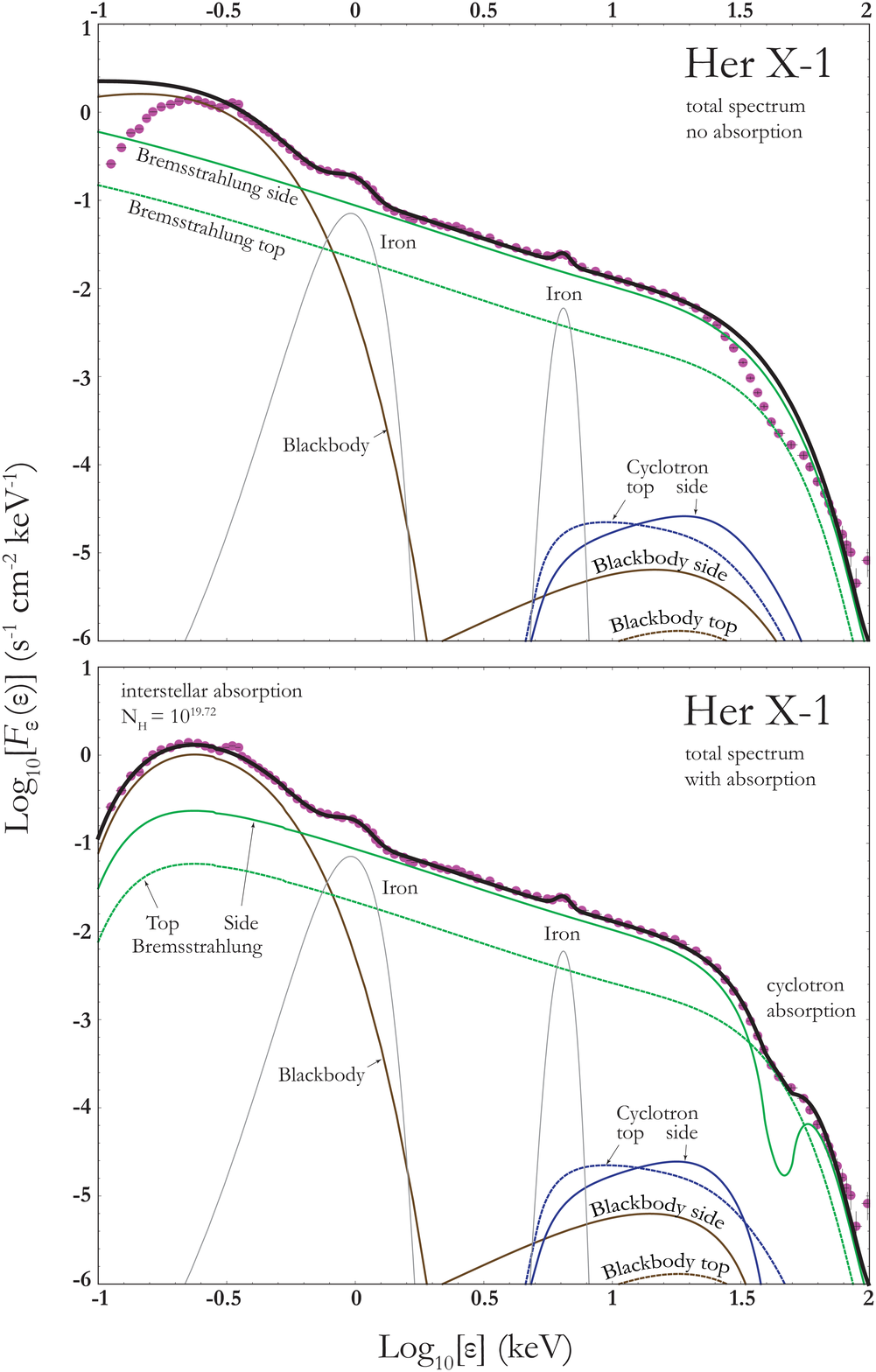}
\caption[Her X-1 Spectrum]{Phase-averaged spectrum for Her X-1 calculated using Equation (\ref{eqn:wall phase-averaged photon count rate spectrum}). The lower panel includes cyclotron and interstellar absorption, and the upper panel neglects absorption.}
\label{fig:Her X-1 phase-averaged spectrum}
\end{figure}

\subsection{Cen X-3}

Our primary results for the dynamical and radiative properties of Cen X-3 are plotted in Figure~\ref{fig:Cen X-3 phase-averaged properties}. Overall, the properties of the bulk fluid and the shape of the radiation patterns are similar to those found in the case of Her X-1. The fluid enters the top of the accretion column at a distance of 14.25\,km from the surface, passes through a radiation sonic surface at 2.21\,km, and then approaches the stellar surface with a residual stagnation velocity $v_* = -0.0081\,c$. The region in the column above the radiation sonic surface is again dominated by Compton scattering, as the photons transfer their energy to the electrons, while below the sonic surface, the electrons tend to re-heat the photons. The peak magnitudes of the advective and diffusive radiation flux components are nearly 50\% larger than the corresponding peaks in the case of Her X-1. The imprint radius of the cyclotron absorption feature, $r_{\rm cyc}$, is located 1.2\,km below the radiation sonic point, and only 940\,m above the stellar surface, and there is $\sim 10\%$ deviation between $r_{\rm cyc}$ and the radius of maximum X-ray emission, $r_{\rm X}$.

The plots of the thermal and bulk Compton $y$-parameters in Figure~\ref{fig:Cen X-3 phase-averaged properties} indicate that bulk Comptonization dominates in the column above the radiation sonic surface, and thermal Comptonization dominates below it. Again, we see that the shape of the spectrum is essentially created in the lower 2\,km of the column, where both $y$-parameters exceed unity. The cyclotron energy at 31.79\,keV is approximately two-thirds the magnitude seen in Her X-1, which helps to explain why the average photon energy is also less, remaining in the range 10.8\,keV to 14.5\,keV.

\begin{figure}[htbp]
\centering
\includegraphics[width=6.0in]{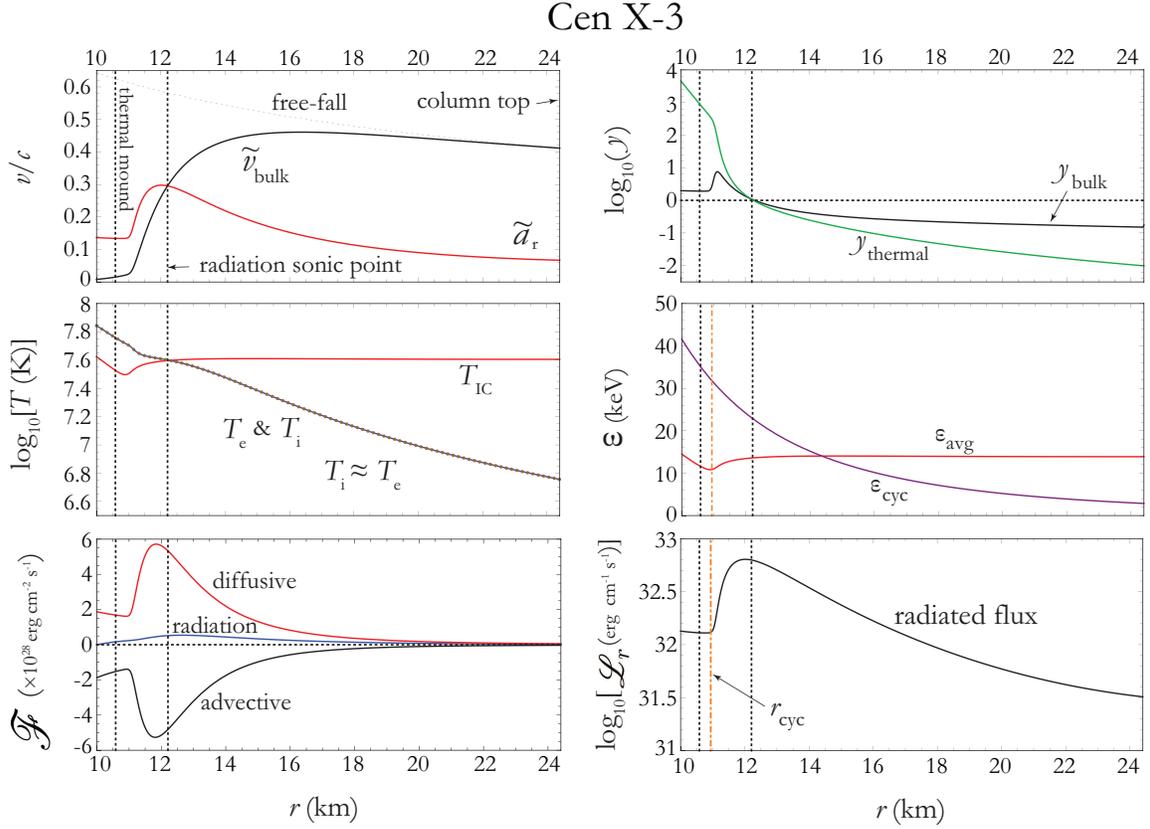}
\caption[Cen X-3 properties]{Same as Figure~\ref{fig:Her X-1 phase-averaged properties}, except here we treat Cen X-3.}
\label{fig:Cen X-3 phase-averaged properties}
\end{figure}

Figure \ref{fig:Cen X-3 radius-dependent and partially-occulted spectra} depicts the radius-dependent and partially-occulted spectra for Cen X-3, at four indicated values for the radius $r$. In the case of the radius-dependent spectra, we plot the emergent spectrum per cm at the precise radius $r$, whereas in the partially-occulted case, the value of $r$ corresponds to the lower bound of integration in Equation~(\ref{eqn:wall phase-averaged photon count rate spectrum}). A qualitative comparison shows that the dominant contribution to the fan-beam component is emitted near the radiation sonic surface, at 2.21\,km above the stellar surface, which is an order of magnitude larger than the contribution radiated at 10.8\,km ($\sim $3/4 column length). In Cen X-3 we observe an extended sinking regime below the thermal mound, where the average photon energy begins to rise. The average photon energy achieves its highest value at the stellar surface, which is not the case with Her X-1. The spectral contribution emitted below the thermal mound is insignificant, demonstrating that most of the photons injected in this region are absorbed before they can escape.

The best visual match of the phase-averaged photon count rate spectrum for Cen X-3 is depicted in Figure \ref{fig:Cen X-3 phase-averaged spectrum}. In the upper panel, we plot the raw, non-attenuated spectrum, and in the lower panel we plot the spectrum with interstellar absorption and cyclotron absorption included. The hydrogen column density is set equal to ${\rm N_H} = 1.58 \times 10^{22}$\,cm$^{-2}$, which agrees well with the value ${\rm N_H} = 1.95 \times 10^{22}$\,cm$^{-2}$ obtained by Burderi et al. (2000). We have simulated the effect of iron emission by adding a Gaussian at 6.67\,keV, with the auxiliary parameters listed in Table~\ref{tab:Auxiliary Parameters}. A side-by-side comparison with the Her X-1 spectrum shows that the cyclotron and blackbody contributions for Cen X-3 are more broad at energies $\lesssim1$keV, and the cyclotron component is strongest at energies below $\lesssim0.3$\,keV. The cyclotron absorption feature is centered at energy 31.79\,keV, which is much lower than the 44.72\,keV feature in Her X-1. The spectrum is not as hard as Her X-1 at the highest energies, which is what we expect because the average photon energy is lower. The overall Cen X-3 emission geometry is dominated by the fan beam component, and the fan-to-pencil ratio is larger than that obtained for Her X-1.

\begin{figure}[htbp]
\centering
\includegraphics[width=5.0in]{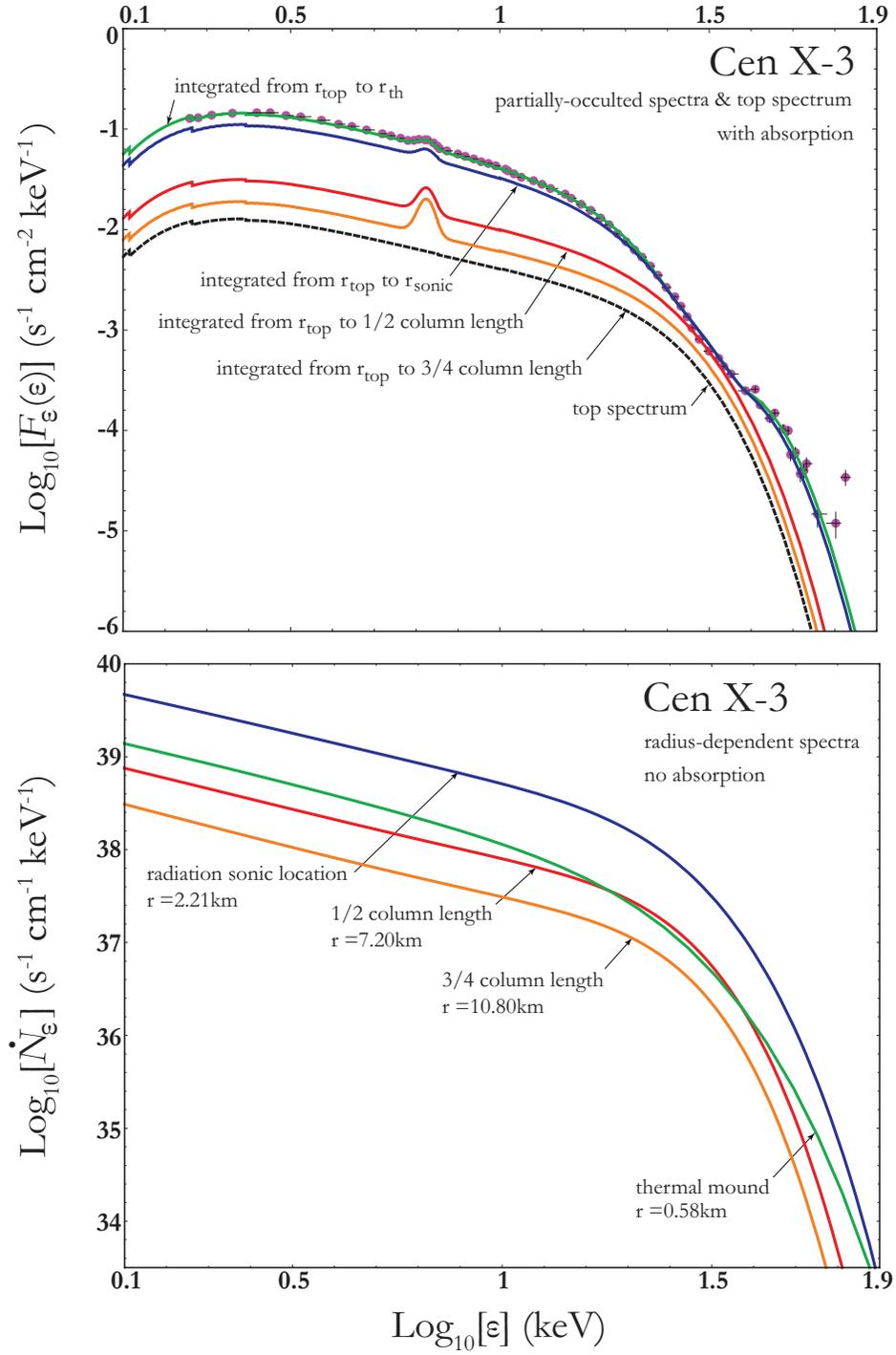}
\caption[Cen X-3 Spectrum]{Same as Figure~\ref{fig:Her X-1 radius-dependent and partially-occulted spectra}, except here we treat Cen X-3.}
\label{fig:Cen X-3 radius-dependent and partially-occulted spectra}
\end{figure}
\begin{figure}[htbp]
\centering
\includegraphics[width=5.0in]{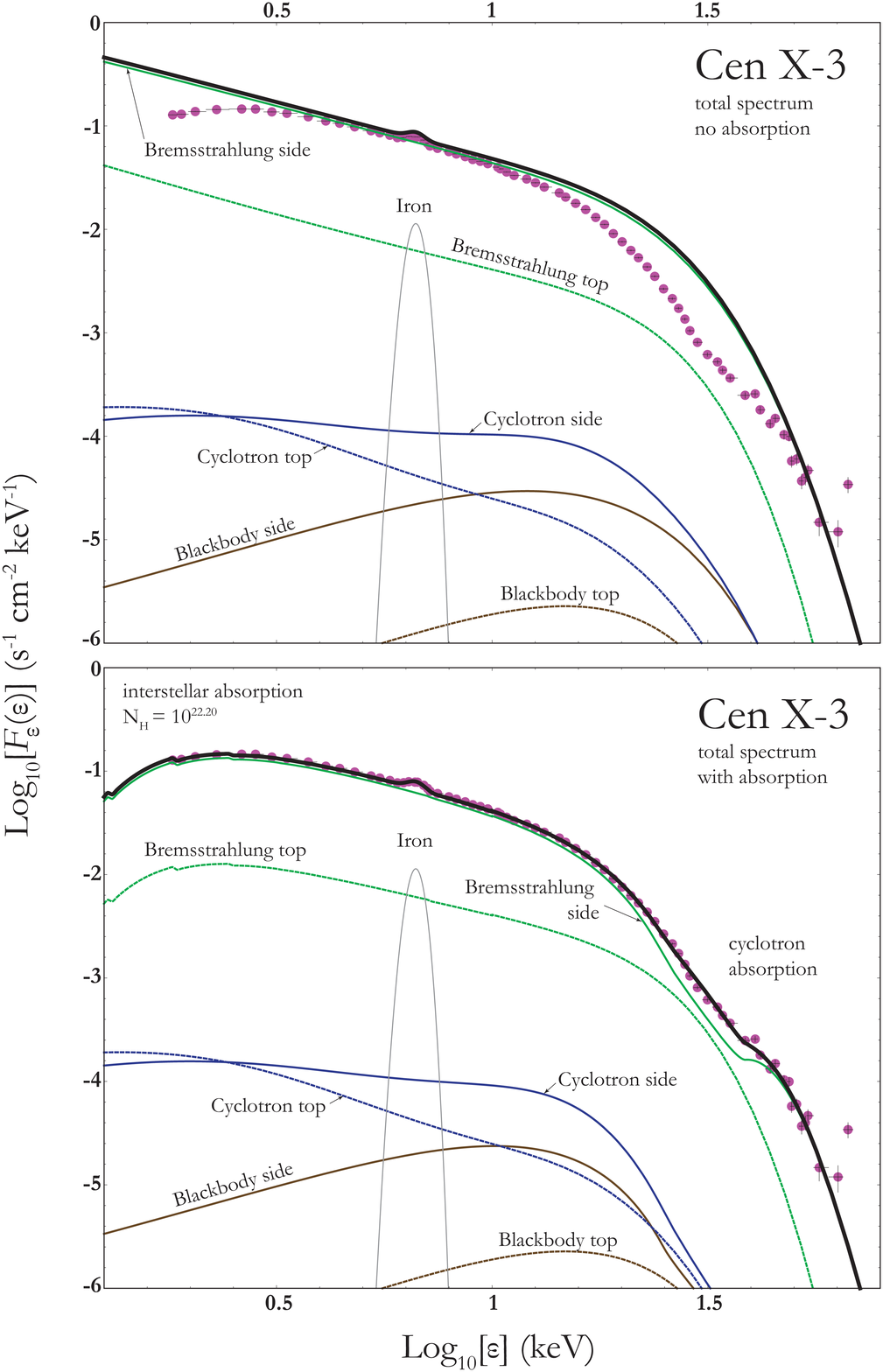}
\caption[Cen X-3 Spectrum]{Same as Figure~\ref{fig:Her X-1 phase-averaged spectrum}, except here we treat Cen X-3.}
\label{fig:Cen X-3 phase-averaged spectrum}
\end{figure}

\subsection{LMC X-4}

The dynamical and radiative properties of LMC X-4 are depicted in Figure \ref{fig:LMC X-4 phase-averaged properties}. The results are similar to those obtained for both Her X-1 and Cen X-3 in many respects, but there are also some important differences. The fluid enters the top of the accretion column at altitude 11.30\,km above the stellar surface, and passes through the radiation sonic surface at radius 13.21\,km. At the stellar surface, the gas has a residual stagnation velocity $v_* = -0.0098\,c$. Comparing the dynamical results for LMC X-4, Her X-1, and Cen X-3, we note that the radius of the radiation sonic surface tends to increase as we move to high-luminosity sources. The extended sinking regime likewise becomes more extended with increasing luminosity. The photon-electron energy exchange in LMC X-4 is similar to that observed for Her X-1 and Cen X-3, with all three temperatures ($T_{\rm IC}(r)$, $T_i(r)$, and $T_e(r)$) exhibiting a slow exponential increase in the sinking regime.

In contrast to Her X-1 and Cen X-3, the imprint radius of the cyclotron absorption feature in LMC X-4 is located 1.05\,km {\it above} the radiation sonic surface, at an altitude of 4.26\,km above the stellar surface, and the cyclotron energy for LMC X-4 is found to be 32.15\,keV. Figure~\ref{fig:LMC X-4 phase-averaged properties} indicates that in LMC X-4, the thermal Compton parameter, $y_{\rm thermal}$, exceeds the value of the bulk Compton parameter, $y_{\rm bulk}$, below about $r \sim 16\,$km in the column, whereas this transition occurs closer to $r \sim 12\,$km in Her X-1 and Cen X-3 (see Figures~\ref{fig:Her X-1 phase-averaged properties} and \ref{fig:Cen X-3 phase-averaged properties}). This explains the flatter slope of the power-law section of the spectrum in the plot of the phase-averaged spectrum for LMC X-4 (see Figure~\ref{fig:LMC X-4 radius-dependent and partially-occulted spectra}). We note that in the case of LMC X-4, the value of the cyclotron imprint radius, $r_{\rm cyc}$, differs from the maximum X-ray emission radius, $r_{\rm X}$, by $\sim 10\%$. 
\begin{figure}[htbp]
\centering
\includegraphics[width=6.0in]{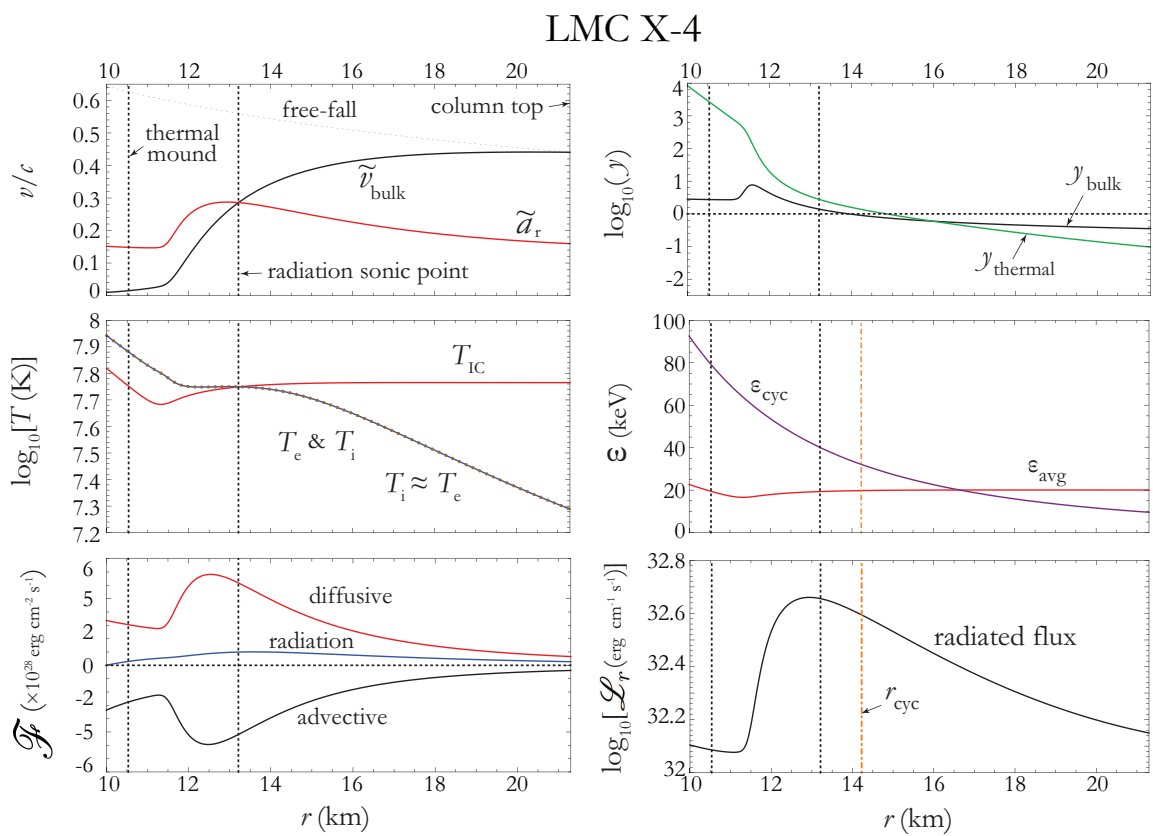}
\caption[LMC X-4 properties]{Same as Figure~\ref{fig:Her X-1 phase-averaged properties}, except here we treat LMC X-4.}
\label{fig:LMC X-4 phase-averaged properties}
\end{figure}

Figure~\ref{fig:LMC X-4 radius-dependent and partially-occulted spectra} depicts the results obtained for the radius-dependent and partially-occulted spectra for LMC X-4. We again observe that the emission is concentrated in the vicinity of the radiation sonic surface. The phase-averaged photon count rate spectrum is plotted in Figure \ref{fig:LMC X-4 phase-averaged spectrum}. The hydrogen column density is set equal to ${\rm N_H} = 9.22 \times 10^{21}$ cm$^{-2}$, which is somewhat larger than previous estimates of $\sim 0.57 \times 10^{21}$\,cm$^{-2}$ (Dickey \& Lockman 1990, La Barbera et al. 2001, Paul et al. 2002). The iron Gaussian is centered at 5.90\,keV (see Table~\ref{tab:Auxiliary Parameters}). The LMC X-4 spectrum is harder than that of Her X-1 and Cen X-3, which explains the larger values obtained for the average photon energy, in the range 16.6\,keV to 22.7\,keV. The emission geometry indicates strong contributions from both fan and pencil beam components in this source, although the fan component still dominates in the phase-averaged spectrum.

\begin{figure}[htbp]
\centering
\includegraphics[width=5.0in]{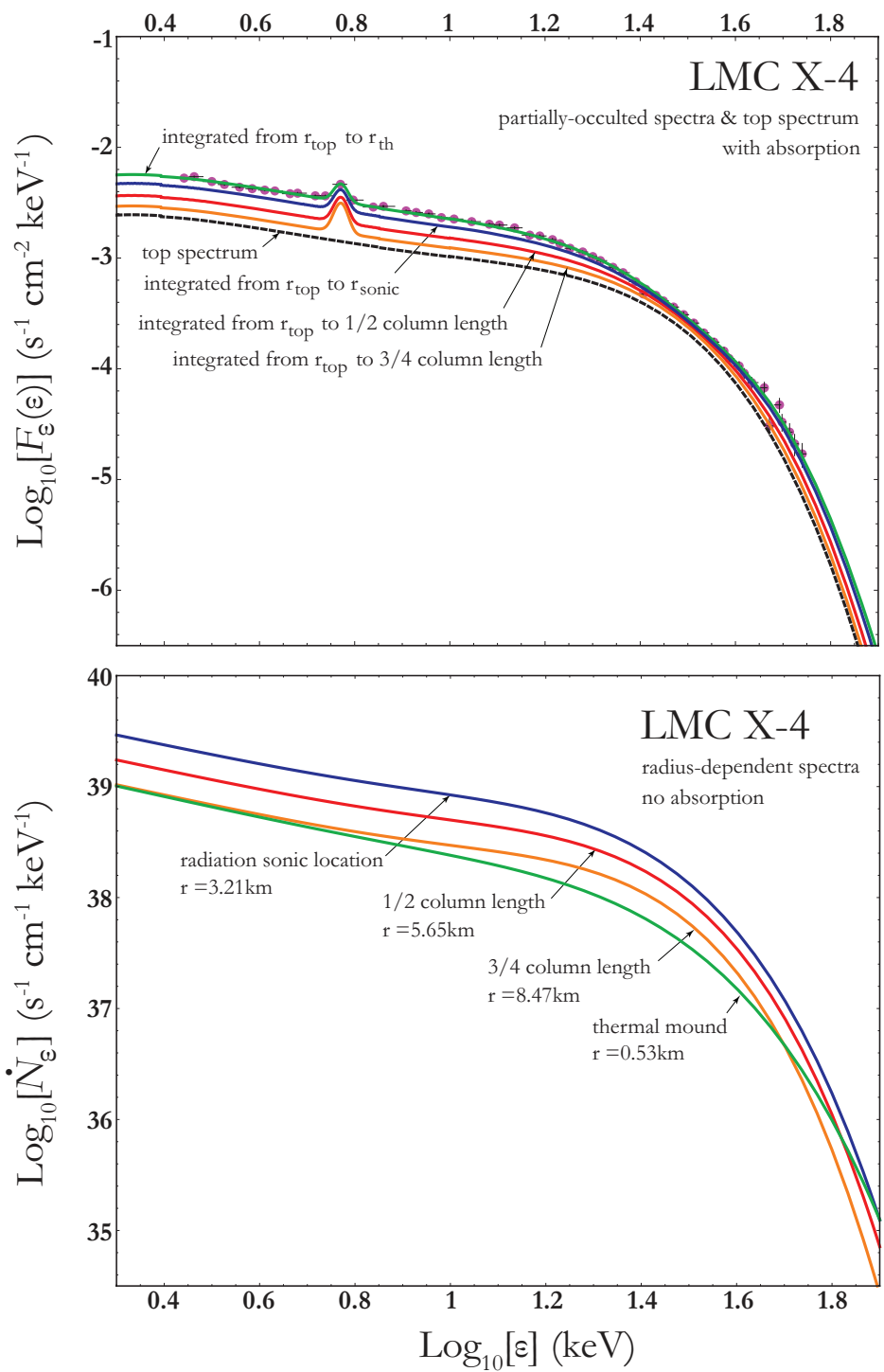}
\caption[LMC X-4 Spectrum]{Same as Figure~\ref{fig:Her X-1 radius-dependent and partially-occulted spectra}, except here we treat LMC X-4.}
\label{fig:LMC X-4 radius-dependent and partially-occulted spectra}
\end{figure}
\begin{figure}[htbp]
\centering
\includegraphics[width=5.0in]{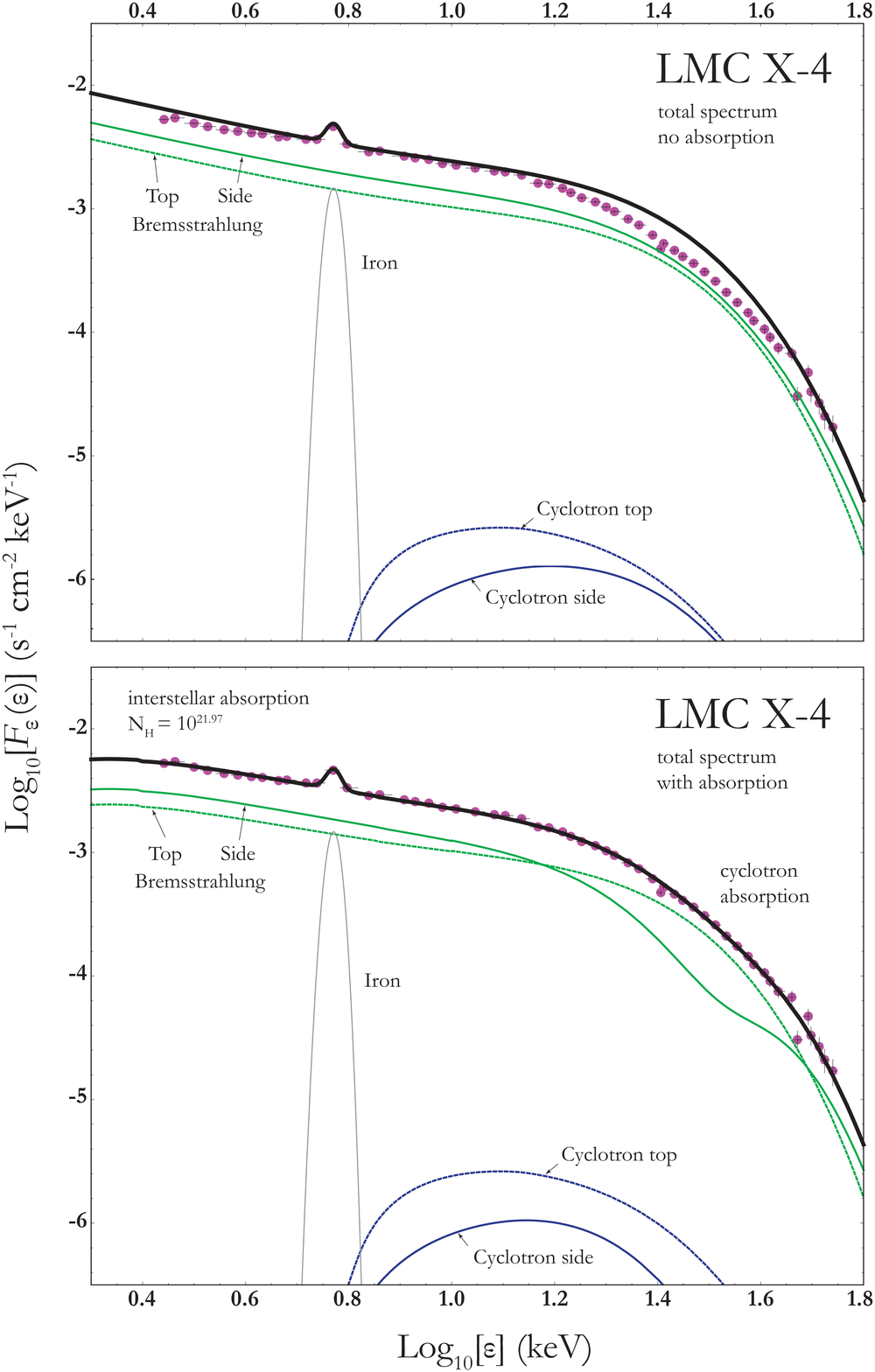}
\caption[LMC X-4 Spectrum]{Same as Figure~\ref{fig:Her X-1 phase-averaged spectrum}, except here we treat LMC X-4.}
\label{fig:LMC X-4 phase-averaged spectrum}
\end{figure}

\section{DISCUSSION}
\label{section:DISCUSSION AND CONCLUSION}

The coupled radiative-hydrodynamical simulation developed here, and the associated calculation of the phase-averaged photon count rate spectrum, represents the first self-consistent model for the hydrodynamical and radiative properties of accretion-powered X-ray pulsars. The model is cast in a realistic dipole geometry, and includes rigorous calculations of the radial profiles of the flow velocity $v(r)$ and the electron temperature $T_e(r)$. We have iteratively solved an ODE-based hydrodynamical code that determines the dynamical structure of the accretion column, coupled with a PDE-based radiative code that computes the radiation spectrum as a function of radius and energy by employing the finite-element method to solve a second order, elliptical, nonlinear partial differential equation. The inverse-Compton temperature profile, $T_{\rm IC}(r)$, provides the link between the hydrodynamics developed in Paper I, and the solution for the radiation distribution function developed here in Paper II. The iterative process converges to yield a self-consistent description of (1) the dynamical structure over the full length of the accretion column, and (2) the energy distribution in the emergent radiation field.

Our model employs a ``mirror'' boundary condition for the radiation field at the stellar surface, which means that we strive to obtain zero net radiation flux there. This condition is satisfied in Cen X-3 and LMC X-4, but in the case of Her X-1, there is a slightly positive residual radiation energy flux at the stellar surface. The average photon energy is relatively constant in all three sources, and is comparable to the cyclotron energy near the cyclotron imprint radius.

Our analysis of the $y$-parameters show that bulk Comptonization dominates thermal Comptonization in the top 80\% of the column for both Her X-1 and Cen X-3, but in LMC X-4, which is also the brightest source of the three pulsars, bulk Comptonization dominates only in the top 50\%. The shape of the spectrum, however, is altered only in the region dominated by thermal Comptonization, when the $y$-parameters exceed unity, and we conclude that the spectrum shape is significantly dependent on thermal Comptonization. This takes place within approximately 2\,km of the stellar surface in the case of Her X-1 and Cen X-3, while for LMC X-4 the region extends to nearly 5\,km above the surface.

The magnitude of the radiated fan component reaches its maximum value near the radiation sonic surface for each source. In the case of Her X-1, the cyclotron absorption imprint radius matches the maximum luminosity location. However, in the cases of Cen X-3 and LMC X-4, the imprint radius is slightly below and above it, respectively. The reason for this offset is not yet known. It is reasonable to argue the absorption imprint radius should correspond to the radius of maximum wall emission. One possible explanation is that our assumed values of stellar mass and radius for Cen X-3 and LMC X-4 require modification, which is planned in future work.

\subsection{Cyclotron Absorption Feature}
\label{sec:cycabsorp}

The Cyclotron Resonant Scattering Feature (CRSF) is a complex shape and often non-Gaussian (Sch\"onherr et al. 2007). A correct treatment of the line profile should include modeling the physical causes for the natural line width and Doppler broadening (Tr\"umper et al. 1977; Harding \& Lai 2006). This is currently a noted limitation in the present state of our model. The cyclotron absorption feature was approximated using a 2D Gaussian function (see Equation (\ref{eqn:2D Gaussian final form})), which mathematically describes an ad-hoc multiplication of two 1D Gaussian functions. We introduced standard deviations in both the energy and spatial dimensions, given by $\sigma_{\epsilon}$ and $\sigma_r$, respectively, and we also introduced a joint strength parameter $d_{cr}$.

The model values obtained for the spatial standard deviation, $\sigma_r$, are 9.20\,km (Her~X-1), 5.17\,km (Cen X-3), and 4.14\,km (LMC X-4). We expect that in the case of Her X-1, the relative width of the CRSF line, $\Delta \epsilon$, can be approximated by $\Delta \epsilon / \epsilon_{\rm cyc}  < 0.35$ (e.g., Staubert et al. 2014). Combining this with the dipole variation of the magnetic field, with $B \propto r^{-3}$, we should observe a relative radial width on the order of $\Delta r_{\rm cyc} / r_{\rm cyc} < 0.1$ (Tr\"umper et al. 1977). Using our Her X-1 result for the cyclotron imprint radius, $r_{\rm cyc} = 11.74\,$km, we would expect that the CRSF-forming region would span a radial domain of width $\Delta r_{\rm cyc} \sim 1.2$\,km, which is much less than the radial standard deviation of 9.2\,km that we have used here. Similar arguments can be made for Cen X-3 and LMC X-4, although not quite as severe. The standard deviation in the energy domain, for all three sources, was very close to $\sigma_{\rm cyc} \sim 11\,$keV, which is still relatively large, but not unreasonable, when compared with values commonly used in XSPEC to model the cyclotron line width. Finally, the joint strength parameter, $d_{cr}$, is purely an attempt to describe the magnitude of the combined effects of the two 1D Gaussian function, with little justification other than to obtain a spectrum best visual match in the CRSF region.

\subsection{Electron Scattering Cross-Section}
\label{sec:elecscat}

The electron parallel scattering cross-section plays a central role in the deceleration of the bulk fluid and the corresponding escape of radiation through the walls. Becker et al. (2012) showed that the dynamical timescale for fluid deceleration and photon escape timescale must be comparable in the sinking regime below the radiation sonic surface. The theoretical value for the parallel scattering cross-section is given in Equation~(17) of Becker et al. (2012), which can be written as
\begin{equation}
L_{\rm X} = \frac{G M_* m_p c}{\sigma_\parallel} \frac{\pi r_0^2}{R_*^2} \ .
\label{eqn:theoretical value of parallel scattering cross-section}
\end{equation}
This relation is applicable to the three sources treated here, which are in the luminosity range $L_{\rm X} \sim 10^{37-38}$\,erg\,s$^{-1}$. In our model, $\sigma_\parallel$ is a fitting parameter, so it is interesting to compare the values we obtained for this parameter with those computed using Equation~(\ref{eqn:theoretical value of parallel scattering cross-section}). This comparison is carried out in Table~\ref{tab:Parallel Scattering Cross-Section}. We note that in the case of Cen X-3, the two values agree to within $\sim14$\% while the error for LMC X-4 is $\sim13$\%. In the case of Her X-1, the error is larger, but the two values still agree to within a factor of $\sim 3$.

\begin{deluxetable}{lccc}
\tablecolumns{4}
\tablewidth{0pt}
\tablecaption{Parallel Scattering Cross-Section
\label{tab:Parallel Scattering Cross-Section}}
\tablehead{
\colhead{Source}
& \multicolumn{1}{p{3cm}}{\centering $\sigma_{\parallel}/\sigma_{\rm T}$ \\ This Paper}
& \multicolumn{1}{p{3.5cm}}{\centering $\sigma_{\parallel}/\sigma_{\rm T}$ \\ Becker et al. 2012}
& \colhead{ratio}
}
\startdata
Her X-1 & $1.02 \times 10^{-3}$ & $3.44 \times 10^{-4}$ & 2.96\\
Cen X-3 & $7.51 \times 10^{-4}$ & $8.78 \times 10^{-4}$ & 0.86\\
LMC X-4 & $4.18 \times 10^{-4}$ & $4.78 \times 10^{-4}$ & 0.87\\
\enddata
\end{deluxetable}

\subsection{Partially-Occulted Spectra}
\label{sec:pospectra}

In Figure \ref{fig:Radius spectra} we plot the radius-dependent (left column) and partially-occulted (right column) spectra for all three sources at four radii: the thermal mound surface (green), radiation sonic surface (blue), and at distances from the stellar surface equal to 1/2 and 3/4 of the column length (red and orange), respectively. For illustration purposes, interstellar absorption and cyclotron absorption are included only in the plots for the partially-occulted spectra. The partially-occulted spectra are interpreted as resulting from the rotation of the pulsar, which causes various portions of the accretion column to disappear behind the neutron star. Although the partially-occulted spectra gives some hints into how the rotation of the star would affect the observed spectrum, we cannot make any definitive conclusions in the absence of a complete model for the stellar rotation and the emission geometry, including the angle between the spin axis and the line of sight to the observer, and also the angle between the spin axis and the radiating magnetic pole (or poles). Such a model should also include the effects of general relativistic light bending, and also the effect of the directional dependence of the cyclotron absorption and emission processes occurring in the outer sheath of the column surface. A calculation of this type is beyond the scope of the present paper, but we plan to pursue the development of a complete geometrical picture in future work.

The results obtained here for the partially-occulted spectra establish that the largest spectral contributions occur near the radiation sonic surface. We find that the observed spectra are essentially generated in the last few kilometers above the stellar surface, where the $y$-parameters exceed unity and thermal Comptonizaton dominates. The sequence of partially-occulted spectra plotted for each source demonstrate that the observed spectra are due to emission escaping from altitudes between the column top and the thermal mound. The escape of photos from altitudes below the top of the thermal mound makes a negligible contribution to the total spectra, due to the dense atmosphere in the extended sinking regime, where the parallel absorption optical depth exceeds unity, and most of the photons are absorbed.

\begin{figure}[htbp]
\centering
\includegraphics[width=6.2in]{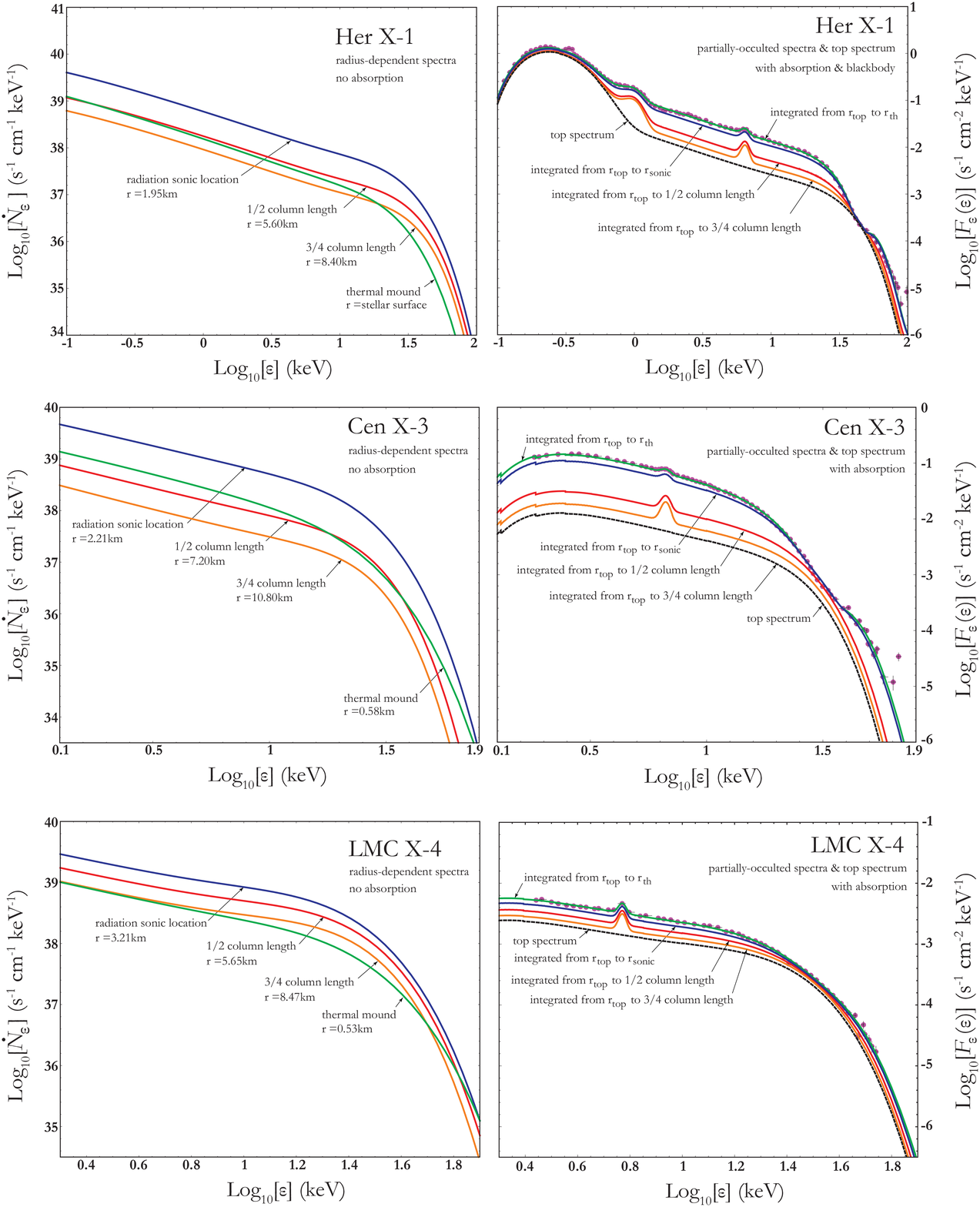}
\caption[Radius spectra]{Radius-dependent (left column) and partially-occulted spectra (right column) computed using Equations (\ref{eqn:Ndot}) and (\ref{eqn:wall phase-averaged photon count rate spectrum}), respectively.}
\label{fig:Radius spectra}
\end{figure}

The two-dimensional approximations employed in the Wang \& Frank (1981) model suggest that $\sim 90$\% of the observed radiation escapes in the fan-beam component, with only $\sim 10$\% contributed by the pencil beam. As these authors note, the principal weakness of their model is the lack of any implementation of energy exchange between the photons and the plasma. Our model results establish that the fan-beam contribution generally dominates over the pencil beam, although this can be violated in the vicinity of the cyclotron absorption energy. We also note that the new model provides additional radius-dependent and energy-dependent information based upon realistic physics within the column, which is unavailable in the model of Wang \& Frank (1981).

Langer \& Rappaport (1982) developed a hydrodynamical model for accretion flows in lower luminosity X-ray pulsars ($L_{\rm X} \lesssim 1.9 \times 10^{36}\,{\rm erg\,s}^{-1}$). Their model includes a partial implementation of the energy exchange processes linking the photons with the plasma, but it does not treat the radiation transfer problem in detail. A major goal of our work has been to properly model the interactions between the gas and the radiation due to thermal emission and absorption, cyclotron emission and absorption, and Compton scattering. Hence our model provides a self-consistent description of the hydrodynamics, the thermodynamics, and the radiative transfer occurring in the accretion column. Our results suggest that, even for low-luminosity sources, the coupling of the radiation with the matter via Compton energy transfer can have a profound effect on the electron temperature profile. Bykov \& Krassilchtchikov (2004) also computed the hydrodynamics of a 1D, two-fluid model in a dipole magnetic field geometry for low-luminosity sources. While they go one step further than Langer \& Rappaport (1982) by adopting a cyclotron diffusion approximation to treat the radiation transport, they do not account for the fundamental effects of bulk and thermal Comptonization, nor do they include bremsstrahlung or blackbody source photons.

We can gain further insight into the relative importance of the pencil- and fan-beam components in forming the observed spectra of the X-ray pulsars treated here. In Figure \ref{fig:ratio spectral flux emission geometry} we plot emission ratio curves, formed by dividing the emitted pencil-beam photon spectrum by the column-integrated (phase-averaged) fan-beam photon spectrum for Her X-1 (blue), Cen X-3 (black), and LMC X-4 (green). In general, for all three sources, the emission from the wall (fan) component exceeds that escaping from the column top (pencil) component, except within a narrow energy range centered on the cyclotron absorption feature. The total observed photon energy and number fluxes associated with the pencil and fan components can be calculated by integrating the respective spectra over the energy band from $\epsilon_{\rm min}$ to $\epsilon_{\rm max}$. The observed total number flux emitted in the fan-beam component, emanating through the column walls, is calculated using
\begin{equation}
\mathscr{F}^{\rm tot}_{\rm ph} = \int_{\epsilon_{\rm min}}^{\epsilon_{\rm max}}
F^{\rm tot}_\epsilon(\epsilon) \, d \epsilon \quad \propto \quad {\rm s^{-1} \, cm^{-2}} \ ,
\label{eqn:photon number flux wall}
\end{equation}
where $F^{\rm tot}_\epsilon(\epsilon)$ is given by Equation (\ref{eqn:phase-averaged photon count rate spectrum}). The corresponding observed total energy flux is given by
\begin{equation}
\mathscr{F}^{\rm{tot}}_{\rm en} = \int_{\epsilon_{\min}}^{\epsilon_{\max}} \epsilon \, F^{\rm{tot}}_{\epsilon}(\epsilon) \, d \epsilon \quad \propto \quad {\rm keV \, s^{-1} \, cm^{-2}} \ .
\label{eqn:photon energy flux wall}
\end{equation}
Likewise, the total observed number flux due to the pencil-beam emission escaping from the top of the column is given by
\begin{equation}
\mathscr{\hat F}^{\rm tot}_{\rm ph} = \int_{\epsilon_{\rm min}}^{\epsilon_{\rm max}}
\hat F^{\rm tot}_\epsilon(\epsilon) \, d \epsilon \quad \propto \quad {\rm s^{-1} \, cm^{-2}} \ ,
\label{eqn:photon number flux top}
\end{equation}
where $\hat F^{\rm tot}_\epsilon(\epsilon)$ is evaluated using Equation (\ref{eqn:top photon count rate spectrum}). The corresponding total energy flux is computed using
\begin{equation}
\mathscr{\hat{F}}^{\rm{tot}}_{\rm en} = \int_{\epsilon_{\min}}^{\epsilon_{\max}} \epsilon \, \hat{F}^{\rm{tot}}_{\epsilon}(\epsilon) \, d \epsilon \quad \propto \quad {\rm keV \, s^{-1} \, cm^{-2}} \ .
\label{eqn:photon energy flux top}
\end{equation}
The four observed fluxes, calculated using Equations (\ref{eqn:photon number flux wall}) - (\ref{eqn:photon energy flux top}), are listed for each source in Table~\ref{tab:Emission Geometry}, separated into the contributions due to the pencil-beam and fan-beam components.} In general, the majority of the energy and number fluxes is contributed by the fan beam, although in the case of LMC X-4, the pencil beam number flux is comparable to the fan beam contribution.

\begin{figure}[htbp]
\centering
\includegraphics[width=4.5in]{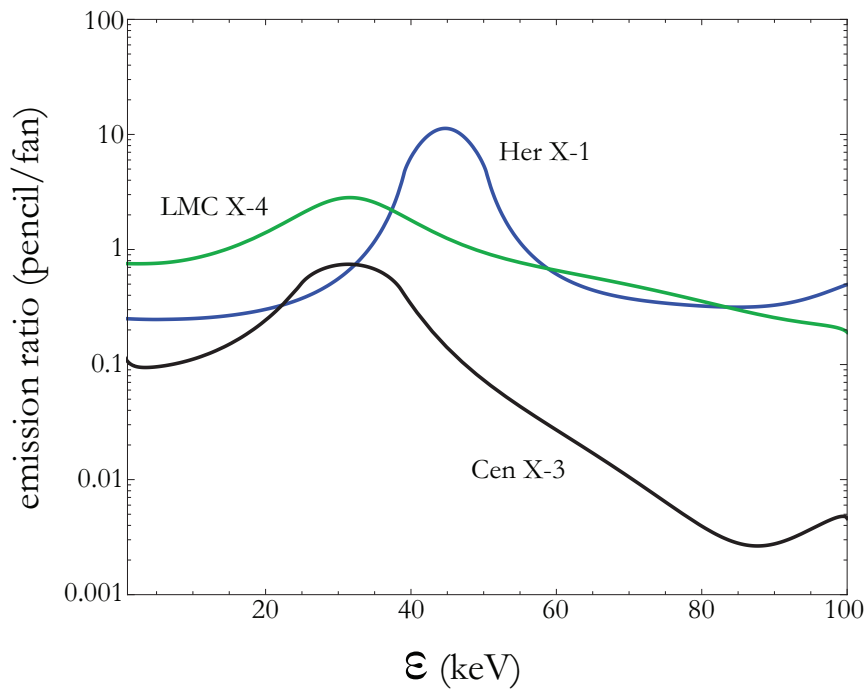}
\caption[Emission geometry]{Ratio of the pencil-beam photon spectrum divided by the fan-beam photon spectrum, plotted as a function of the photon energy $\epsilon$. See the discussion in the text.}
\label{fig:ratio spectral flux emission geometry}
\end{figure}

\begin{deluxetable}{lcccc}
\tablecolumns{4}
\tablewidth{0pt}
\tablecaption{Total Observed Photon Number and Energy Fluxes
\label{tab:Emission Geometry}}
\tablehead{
\colhead{Source}
& \colhead{Pencil (Top)}
& \colhead{Fan (Wall)}
& \colhead{ratio}
}
\startdata
Number Flux \; ($\propto \; \rm{cm^{-2} \, sec^{-1}}$) & $\mathscr{\hat F}^{\rm tot}_{\rm ph}$ & $\mathscr{F}^{\rm tot}_{\rm ph}$ & Top/Wall \\
Her X-1 & 0.15 & 0.57 & 0.27\\
Cen X-3 & 0.40 & 3.07 & 0.13\\
LMC X-4 & 0.05 & 0.06 & 0.85\\
\hline
Energy Flux \; ($\propto \; \rm{keV \, cm^{-2} \, sec^{-1}}$) & $\mathscr{\hat F}^{\rm tot}_{\rm en}$ & $\mathscr{F}^{\rm tot}_{\rm en}$ & Top/Wall \\
Her X-1 & 1.14 & 2.99 & 0.38\\
Cen X-3 & 1.13 & 7.77 & 0.15\\
LMC X-4 & 0.38 & 0.31 & 1.20\\
\enddata
\end{deluxetable}

\subsection{Future Work}
\label{sec:futurework}

Paper I includes a discussion of future enhancements to the hydrodynamical model, which could lead to an improved description of the structure of the accretion column. However, there are also numerous modifications that would facilitate a more detailed investigation of the partially-occulted and phase-averaged spectra and associated spectral properties. For example, as mentioned earlier, our preliminary results for Cen X-3 and LMC X-4 suggest that modification of the stellar mass and radius may improve the agreement between the cyclotron absorption imprint radius, $r_{\rm cyc}$, and the radius at which the luminosity per unit length ($\mathscr{L}_r$) is maximized, denoted by $r_{\rm X}$. In principle, we would expect these two radii to agree.

Changes in the CRSF line energy and X-ray continuum can be studied by varying the accretion rate, $\dot M$, over a specific range, which may provide new insights into the observed correlation between the cyclotron line energy and the X-ray luminosity (Mihara et al. 1995; Staubert et al. 2007; Becker et al. 2012; Rothschild et al. 2016). To obtain a fully self-consistent and detailed theoretical study of the line width and depth, which can be compared with recent observational studies, such as that by Rothschild et al. (2016), we would need to replace the 2D cyclotron Gaussian with a true energy-dependent cyclotron absorption term in both the hydrodynamic and photon transport equations. This should facilitate calculation of the X-ray power-law continuum for any accretion rate, resulting in correlative studies of the hardness ratio with variations of the source luminosity (Postnov et al. 2015).

We have assumed in our model that the dipole magnetic field component dominates over the full length of the accretion column. However, the possible presence of a magnetic quadrupole component may significantly influence the overall field geometry (Shakura et al. 1991; Panchenko \& Postnov 1994), which could substantially modify the locations and shapes of the primary emission regions, and the accompanying emission characteristics (Jardine et al. 2006; Gregory et al. 2006; Long et al. 2007; Donati et al. 2007b; Long et al. 2008). For example, if the quadrupole component is aligned with the dipole, then the accreting matter may impact the surface as a ring rather than a disk (Postnov et al. 2013). This particular possibility is included implicitly in our model, at least qualitatively, since we allow for a partially-filled column, in which case the matter impacts the surface in a ring configuration. However, if the quadrupole component is misaligned with the dipole component, then the accretion structure is likely to be much more complex and asymmetric. The treatment of that type of asymmetric structure would require the development of a two-dimensional spatial model, which is beyond the scope of the work presented here, but it could possibly be developed in the future.

The results we have obtained for the partially-occulted spectra plotted in Figures~\ref{fig:Her X-1 radius-dependent and partially-occulted spectra}, \ref{fig:Cen X-3 radius-dependent and partially-occulted spectra}, and \ref{fig:LMC X-4 radius-dependent and partially-occulted spectra} provide motivation for future studies employing a more realistic source geometry, including the possibility of a quadrupole magnetic field component, which may alter the pulse profiles, as discussed in the context of Her X-1 by Postnov et al. (2013). Here, we have simply explored the consequences of varying the lower integration bound in Equation~(\ref{eqn:wall phase-averaged photon count rate spectrum}) for the observed fan-beam flux component $F_\epsilon(\epsilon)$ as a means to roughly approximate the effect of the stellar rotation, which would cause varying fractions of the lower portion of the column to disappear from view behind the star. The results we have obtained suggest that a careful consideration of the source geometry, relativistic effects, and the directional dependence of the cyclotron scattering cross section should be implemented in future studies.

\appendix

\newpage

\newpage
\section{PHOTON ESCAPE FORMALISM}
\label{section:PHOTON ESCAPE FORMALISM}

In general, the transport equation governing the evolution of the isotropic (angle-averaged) radiation distribution $f$ inside the accretion column can be written in the vector form (see Equation~(\ref{eqn:PDEalt1}))
\begin{eqnarray}
\frac{\partial f}{\partial t} + \vec v \cdot \vec\nabla f
= \frac{n_e \bar\sigma c}{m_e c^2}\frac{1}{\epsilon^2}\frac{\partial}{\partial \epsilon}
\left[\epsilon^4 \left(f + k T_e \frac{\partial f}{\partial \epsilon}\right) \right]
+ \vec\nabla \cdot \left(\kappa \vec\nabla f \right) \nonumber \\
+ \frac{1}{3} (\vec\nabla\cdot\vec v) \, \epsilon
\frac{\partial f}{\partial\epsilon} + \dot f_{\rm prod} + \dot f_{\rm abs} \ .
\label{eqn:A1}
\end{eqnarray}
Assuming azimuthal symmetry around the magnetic field axis, and averaging over the cross-sectional area of the accretion column at radius $r$, we obtain (see Equation~(\ref{eqn:PDEalt2}))
\begin{eqnarray}
\frac{\partial f}{\partial t} + v \frac{\partial f}{\partial r}
= \frac{n_e \bar\sigma c}{m_e c^2}\frac{1}{\epsilon^2}\frac{\partial}{\partial \epsilon}
\left[\epsilon^4 \left(f + k T_e \frac{\partial f}{\partial \epsilon}\right) \right]
+ \frac{1}{A(r)}\frac{\partial}{\partial r}\left[A(r) \kappa \frac{\partial f}{\partial r} \right] \nonumber \\
+ \frac{1}{3 A(r)} \frac{\partial[A(r) v]}{\partial r} \, \epsilon \frac{\partial f}{\partial\epsilon}
+ \dot f_{\rm prod} + \dot f_{\rm abs} + \dot f_{\rm esc} \ ,
\label{eqn:A2}
\end{eqnarray}
where the area of the accretion column, $A(r)$, is given by Equation~(\ref{eqn:area}), and the term $\dot f_{\rm esc}$ represents the diffusive escape of radiation through the walls of the column.

Our goal here is to derive a suitable mathematical form for $\dot f_{\rm esc}$ so that it properly accounts for the free-streaming escape of radiation through the sides of the accretion column. In particular, we wish to derive an expression for the mean escape timescale, $t_{\rm esc}$, such that the escape term in Equation~(\ref{eqn:A2}) can be written in the form
\begin{equation}
\dot f_{\rm esc} = - \frac{f}{t_{\rm esc}} \ .
\label{eqn:A3}
\end{equation}
In order to proceed, we return to Equation~(\ref{eqn:A1}) and expand the vector components of the spatial diffusion term, which is the term of interest here. For our purposes, it is sufficient to adopt cylindrical coordinates, so that we obtain
\begin{equation}
\frac{\partial f}{\partial t}\bigg|_{\rm diff}
\equiv \vec\nabla\cdot\left(\kappa\vec\nabla f \right)
= \frac{1}{\rho}\frac{\partial}{\partial \rho}\left(\rho \, \kappa \frac{\partial f}{\partial\rho}\right)
+ \frac{1}{\rho}\frac{\partial}{\partial \varphi}\left(\frac{\kappa}{\rho} \, \frac{\partial f}{\partial\varphi}\right)
+ \frac{\partial}{\partial z}\left(\kappa \frac{\partial f}{\partial z}\right) \ .
\label{eqn:A4}
\end{equation}
Since we are interested in treating the escape of radiation through the column walls, we will focus on the radial ($\hat\rho$) component of the photon flux. Furthermore, we will assume symmetry with respect to the azimuthal angle $\varphi$, so that $\partial f/\partial\varphi=0$. In this case, the radial component of the diffusion transport in Equation~(\ref{eqn:A4}) reduces to
\begin{equation}
\frac{\partial f}{\partial t}\bigg|_{\rm diff,\,\rho}
= \frac{1}{\rho}\frac{\partial}{\partial \rho}\left(\rho \, \kappa \frac{\partial f}{\partial \rho}\right) \ .
\label{eqn:A5}
\end{equation}

If we suppose for simplicity that the seed radiation is injected at the center of the column ($\rho=0$) and diffuses radially outward to the column surface at radius $\rho=\rho_0$, then the radial transport rate of the photons is conserved. In this case, we can rewrite Equation~(\ref{eqn:A5}) as
\begin{equation}
- \rho \, \kappa \frac{\partial f}{\partial \rho} = C_0 = {\rm constant} \ .
\label{eqn:A6}
\end{equation}
The cross-sectional variation of the spatial diffusion coefficient, $\kappa$, is not known with any certainty, but for our purposes here, it is sufficient to assume that it is given by the power-law form
\begin{equation}
\kappa(\rho) = \kappa_0 \left(\frac{\rho}{\rho_0}\right)^\alpha \ ,
\label{eqn:A6b}
\end{equation}
where $\alpha$ is a constant and $\kappa_0$ denotes the surface value of $\kappa$. Integration of Equation~(\ref{eqn:A6}) with respect to $\rho$ then yields for $f(\rho)$ the solution
\begin{equation}
f(\rho) = f_0 \left[1 - \frac{1}{\alpha} + \frac{1}{\alpha} \left(\frac{\rho}{\rho_0}\right)^{-\alpha}\right] \ ,
\label{eqn:A7}
\end{equation}
where $f_0$ is the surface value of $f(\rho)$, given by
\begin{equation}
f_0 = \frac{C_0}{\kappa_0} \ .
\label{eqn:A8}
\end{equation}

The diffusion approximation employed inside the column must transition to radiation free-streaming at the column surface, and therefore the surface boundary condition for Equation~(\ref{eqn:A6}) can be written as
\begin{equation}
- \kappa \frac{\partial f}{\partial\rho} \bigg|_{\rho_0} = c f_0 \ .
\label{eqn:A9}
\end{equation}
By comparing Equations~(\ref{eqn:A6}) and (\ref{eqn:A9}), we find that the constant $C_0$ is given by
\begin{equation}
C_0 = \rho_0 \, c f_0 \ ,
\label{eqn:A10}
\end{equation}
which can be combined with Equation~(\ref{eqn:A8}) to show that
\begin{equation}
\kappa_0 = \rho_0 \, c \ .
\label{eqn:A11}
\end{equation}

The {\it average} rate of change of the distribution function $f$ due to the diffusive escape of radiation through the column walls is given by the volume-weighted integral
\begin{eqnarray}
\left< \frac{\partial f}{\partial t} \right> \bigg|_{\rm diff,\,\rho} & = & \frac{1}{\pi \rho_0^2} \int_0^{\rho_0}
2 \pi \rho \left(\frac{\partial f}{\partial t} \bigg|_{\rm diff,\,\rho}\right) \, d\rho \nonumber \\
& = & \frac{2}{\rho_0^2} \int_0^{\rho_0} \frac{\partial}{\partial\rho}
\left(\rho \, \kappa \frac{\partial f}{\partial\rho}\right) \, d\rho \label{eqn:A12} \\
& = & \frac{2}{\rho_0^2} \left(\rho \, \kappa \frac{\partial f}{\partial\rho}\right) \bigg|_0^{\rho_0} \nonumber \ .
\end{eqnarray}
Combining the final relation with the surface boundary condition given by Equation~(\ref{eqn:A9}), we find that the average escape rate is given by
\begin{align}
\left< \frac{\partial f}{\partial t} \right> \bigg|_{\rm diff,\,\rho}
= - \frac{2 \, c f_0}{\rho_0} \ .
\label{eqn:A13}
\end{align}
We note that this is nearly identical to the form of Equation~(\ref{eqn:A3}) for $\dot f_{\rm esc}$.

In order to make the correspondence complete, and obtain an expression for the mean escape timescale, $t_{\rm esc}$, we write
\begin{equation}
\left< \frac{\partial f}{\partial t} \right> \bigg|_{\rm diff,\,\rho}
= - \frac{\left<f\right>}{t_{\rm esc}} \ ,
\label{eqn:A14}
\end{equation}
where $\left<f\right>$ denotes the volume-weighted average of $f$, given by the integral
\begin{equation}
\left< f \right> = \frac{1}{\pi \rho_0^2} \int_0^{\rho_0}
2 \pi \rho \, f(\rho) \, d\rho \ .
\label{eqn:A15}
\end{equation}
Using Equation~(\ref{eqn:A7}) to substitute for $f(\rho)$ in Equation~(\ref{eqn:A15}) and carrying out the integration yields
\begin{equation}
\left< f \right> = \left(\frac{3 - \alpha}{2 - \alpha}\right) f_0 \ ,
\label{eqn:A16}
\end{equation}
which can be combined with Equations~(\ref{eqn:A11}) and (\ref{eqn:A13}) to obtain the result
\begin{equation}
\left< \frac{\partial f}{\partial t} \right> \bigg|_{\rm diff,\,\rho}
= - \frac{2 \kappa_0}{\rho_0^2} \left(\frac{2-\alpha}{3-\alpha}\right) \left<f\right> \ .
\label{eqn:A17}
\end{equation}

Comparing this result with Equation~(\ref{eqn:A14}), we find that the mean escape timescale, $t_{\rm esc}$, is given by
\begin{equation}
t_{\rm esc} = \frac{\rho_0^2}{2 \kappa_0} \left(\frac{3-\alpha}{2-\alpha}\right) \ .
\label{eqn:A14b}
\end{equation}
We can substitute for $\kappa_0$ using
\begin{equation}
\kappa_0 = \frac{c \, \ell_0}{3} \ ,
\label{eqn:A14c}
\end{equation}
where $\ell_0$ is the scattering mean-free path at the column surface. The result obtained is
\begin{equation}
t_{\rm esc} = \frac{3 \, \rho_0^2}{2 \, c \, \ell_0} \left(\frac{3-\alpha}{2-\alpha}\right) \ .
\label{eqn:A14d}
\end{equation}
The diffusion velocity for photons propagating perpendicular to the axis, $w_\perp$, can be approximated by writing
\begin{equation}
w_\perp \sim \frac{c}{\rho_0 / \ell_0} \ ,
\label{eqn:A14e}
\end{equation}
and therefore we find that
\begin{equation}
t_{\rm esc} \sim \frac{3 \, \rho_0}{2 \, w_\perp} \left(\frac{3-\alpha}{2-\alpha}\right) \ .
\label{eqn:A14f}
\end{equation}
In the context of our dipole model for the accretion column, it follows that we can obtain an adequate approximation for the escape timescale by writing (cf. Equation~(\ref{eqn:escape time}))
\begin{equation}
t_{\rm esc}(r) = \frac{\ell_{\rm esc}(r)}{w_\perp(r)} \ ,
\label{eqn:A21}
\end{equation}
where $\ell_{\rm esc}(r)$ is the width of the (either hollow or filled) column at radius $r$, and $w_\perp(r)$ is computed using Equation~(\ref{eqn:diffusion velocity}).

\newpage

\section{PHOTON NUMBER AND ENERGY DENSITY EQUATIONS}
\label{section:PHOTON ENERGY DENSITY ODE}

We can derive the conservation equations satisfied by the radiation number density, $n_r$, and energy density, $U_r$, by starting with the transport Equation~(\ref{eqn:PDEalt2}) for the isotopic (angle-averaged) photon distribution function, $f$, which can be rewritten in the form
\begin{eqnarray}
\frac{\partial f}{\partial t} + v \frac{\partial f}{\partial r}
= \frac{n_e \bar\sigma c}{m_e c^2}\frac{1}{\epsilon^2}\frac{\partial}{\partial \epsilon}
\left[\epsilon^4 \left(f + k T_e \frac{\partial f}{\partial \epsilon}\right) \right]
+ \frac{1}{A(r)}\frac{\partial}{\partial r}\left[A(r) \kappa \frac{\partial f}{\partial r} \right] \nonumber \\
+ \frac{1}{3 A(r)} \frac{\partial[A(r) v]}{\partial r} \, \epsilon \frac{\partial f}{\partial\epsilon}
+ \frac{Q_{\rm prod}}{r^2 \Omega} - c \, \alpha_{\rm R} f - \frac{f}{t_{\rm esc}} \ ,
\label{eqn:B1}
\end{eqnarray}
where $A(r)$ denotes the area of the dipole column at radius $r$ (Equation~(\ref{eqn:area})), $t_{\rm esc}$ is the mean residence time (Equation~(\ref{eqn:escape time})), $Q_{\rm prod}$ represents the seed photon sources (Equation~(\ref{eqn:fdot sources term})), and $\alpha_{\rm R}$ is the Rosseland mean absorption coefficient (Equation~(\ref{eqn:rosseland})).

Based on Equation~(\ref{eqn:B1}), we can obtain a partial differential equation satisfied by the photon energy moment, $I_n$, which is computed from the distribution function $f$ using the integral
\begin{equation}
I_n \equiv \int_0^\infty \epsilon^n f \, d\epsilon \ .
\label{eqn:B2}
\end{equation}
Special cases of $I_n$ are the number and energy densities, given by
\begin{equation}
n_r = I_2 = \int_0^\infty \epsilon^2 f \, d\epsilon \ , \qquad
U_r = I_3 = \int_0^\infty \epsilon^3 f \, d\epsilon \ .
\label{eqn:B3}
\end{equation}
Operating on Equation~(\ref{eqn:B1}) with $\int_0^\infty \epsilon^n \, d\epsilon$ and integrating by parts once, the result obtained is
\begin{eqnarray}
\frac{\partial I_n}{\partial t} + v \frac{\partial I_n}{\partial r}
= \frac{n_e \bar\sigma c}{m_e c^2}\left\{
\epsilon^{n+2}\left(f + k T_e \frac{\partial f}{\partial \epsilon}\right) \bigg|_0^\infty
- (n-2) \int_0^\infty \epsilon^{n+1} \left(f + k T_e \frac{\partial f}{\partial \epsilon}\right)
d\epsilon
\right\} \nonumber \\
+ \frac{1}{3 A(r)} \frac{\partial[A(r) v]}{\partial r} \left[\epsilon^{n+1} f \bigg|_0^\infty
- (n+1) \int_0^\infty
\epsilon^n f d\epsilon\right] \label{eqn:B4} \\
+ \frac{1}{A(r)}\frac{\partial}{\partial r}\left[A(r) \kappa \frac{\partial I_n}{\partial r} \right]
+ \dot I_n^{\rm prod} - c \, \alpha_{\rm R} I_n - \frac{I_n}{t_{\rm esc}} \ , \nonumber
\end{eqnarray}
where
\begin{equation}
\dot I_n^{\rm prod} \equiv \int_0^\infty \epsilon^n \, \frac{Q_{\rm prod}}{r^2 \Omega} \, d\epsilon \ .
\label{eqn:B5}
\end{equation}

In order to ensure that a finite value is obtained for the radiation energy density, $U_r$, we are required to apply boundary conditions on $f$ such that
\begin{equation}
\lim_{\epsilon \to 0} \ \epsilon^4 f = 0 \ , \qquad
\lim_{\epsilon \to \infty} \ \epsilon^4 f = 0 \ .
\label{eqn:B6}
\end{equation}
Likewise, in order to obtain a finite value for the radiation number density, $n_r$, we must ensure that $f$ satisfies the boundary conditions
\begin{equation}
\lim_{\epsilon \to 0} \ \epsilon^3 f = 0 \ , \qquad
\lim_{\epsilon \to \infty} \ \epsilon^3 f = 0 \ .
\label{eqn:B7}
\end{equation}
The first term on the-right-hand side of Equation~(\ref{eqn:B4}) vanishes once the boundary conditions in Equations~(\ref{eqn:B6}) and (\ref{eqn:B7}) are applied. Next, we integrate by parts again to obtain
\begin{eqnarray}
\frac{\partial I_n}{\partial t} + v \frac{\partial I_n}{\partial r}
= \frac{n_e \bar\sigma c}{m_e c^2}\left\{
- (n-2) \left[I_{n+1} + k T_e \, \epsilon^{n+1} f \bigg|_0^\infty - (n+1) k T_e \int_0^\infty \epsilon^n
f d\epsilon \right] \right\} \nonumber \\
- \left(\frac{n+1}{3}\right) \frac{1}{A(r)} \frac{\partial[A(r) v]}{\partial r} \, I_n
+ \frac{1}{A(r)}\frac{\partial}{\partial r}\left[A(r) \kappa \frac{\partial I_n}{\partial r} \right]
+ \dot I_n^{\rm prod} - c \, \alpha_{\rm R} I_n - \frac{I_n}{t_{\rm esc}} \ .
\label{eqn:B8}
\end{eqnarray}
Applying the boundary conditions in Equations~(\ref{eqn:B6}) and (\ref{eqn:B7}) again, we find that
\begin{eqnarray}
\frac{\partial I_n}{\partial t} + v \frac{\partial I_n}{\partial r}
= \frac{n_e \bar\sigma c}{m_e c^2}\left\{
- (n-2) \left[I_{n+1} - (n+1) k T_e I_n \right]
\right\}
- \left(\frac{n+1}{3}\right) \frac{1}{A(r)} \frac{\partial[A(r) v]}{\partial r} \, I_n \nonumber \\
+ \frac{1}{A(r)}\frac{\partial}{\partial r}\left[A(r) \kappa \frac{\partial I_n}{\partial r} \right]
+ \dot I_n^{\rm prod} - c \, \alpha_{\rm R} I_n - \frac{I_n}{t_{\rm esc}} \ .
\label{eqn:B9}
\end{eqnarray}
This expression can be used to obtain individual conservation equations for the radiation number and energy densities, as discussed below.

By setting $n=2$ in Equation~(\ref{eqn:B9}), we find that the conservation equation for the radiation number density, $n_r$, can be written in the flux-conservation form
\begin{eqnarray}
\frac{\partial n_r}{\partial t}
= - \frac{1}{A(r)} \frac{\partial}{\partial r} \left[A(r) \left(v n_r
- \kappa \frac{\partial n_r}{\partial r} \right) \right]
+ \dot I_2^{\rm prod} - c \, \alpha_{\rm R} n_r - \frac{n_r}{t_{\rm esc}} \ .
\label{eqn:B10}
\end{eqnarray}
Likewise, we can set $n=3$ in Equation~(\ref{eqn:B9}) to obtain a flux-conservation equation for the radiation number density, $U_r$, which can be written as
\begin{eqnarray}
\frac{\partial U_r}{\partial t}
= \frac{n_e \bar\sigma c}{m_e c^2}\left(
1-\frac{T_{\rm IC}}{T_e} \right) 4 k T_e U_r
- \frac{1}{A(r)} \frac{\partial}{\partial r} \left[A(r) \left(\frac{4}{3} v U_r
- \kappa \frac{\partial U_r}{\partial r} \right) \right] \nonumber \\
+ \dot I_3^{\rm prod} - c \, \alpha_{\rm R} U_r - \frac{U_r}{t_{\rm esc}}
+ \frac{v}{3} \frac{\partial U_r}{\partial r} \ ,
\label{eqn:B11}
\end{eqnarray}
where the final term on the right-hand side represents the differential work performed on the radiation by the converging gas in the accretion flow.

{}

\label{lastpage}

\end{document}